\documentclass[a4paper]{jpconf}
\usepackage{graphicx}

\newcommand{\met}{{E}_{T}\!\!\!\!\!\!\slash\,\,\,}
\newcommand{\chin} {\tilde{\chi}_1^0}
\newcommand{\chip} {\tilde{\chi}_1^\pm}
\newcommand{\st} {\tilde{t}}

\begin{document}
\title{Searches for Physics Beyond the Standard Model at Colliders}

\author{Beate Heinemann}

\address{University of California, 305 LeConte Hall, Berkeley, CA 94720 and\\ 
Lawrence Berkeley National Laboratory, MS 50B-6222, Berkeley, CA 94720}

\ead{bheheinemann@lbl.gov}

\begin{abstract}
All experimental measurements of particle physics today are beautifully 
described by the Standard Model. However, there are good reasons to 
believe that 
new physics may be just around the corner at the TeV energy scale. This
energy range is currently probed by the Tevatron and HERA accelerators
and selected results of searches for physics beyond the Standard Model are
presented here. No signals for new physics have been found and limits
are placed on the allowed parameter space for a variety of different particles.
\end{abstract}

\section{Introduction}
Despite the amazing success of the Standard Model (SM) there are strong theoretical motivations
for the existence of new particles could be produced at TeV scale colliders
today or in the very near Future. Many new hypothetical particles have been suggested as a result
of a variety of theoretical models that try to solve the Standard
Model's problems of extrapolating to high energies near the Planck scale.
In this article the recent searches for many different kinds of particles 
are presented: supersymmetric partners of the SM particles, new gauge bosons,
extra spatial dimensions and compositeness. The colliders that presented new results
probing the existence of new particles via direct production 
are the currently operating Tevatron and the very recently closed HERA colliders. 

\section{The Tevatron $p\bar{p}$ and the HERA $e^\pm p$ Colliders}
At the Tevatron in the US near Chicago protons are collided with 
anti-protons with a center-of-mass energy of $\sqrt{s}=1.96$~TeV. The
``Run II'' period of the Tevatron is in progress since 2001 and by now
about 2~fb$^{-1}$ of integrated luminosity have been accumulated by the 
CDF and D0 experiments. The Tevatron is currently the highest energy 
collider world-wide until the startup of the Large Hadron Collider (LHC) 
in Summer 2008.

The HERA $e^\pm p$ collider has operated for fifteen years, between 1992 
and June 2007. While its primary achievement has been its amazing
measurements of the proton structure functions, in particular at low $x$, 
the H1 and ZEUS experiments have also performed many searches for new 
particles some of which will be covered in this article. The total
integrated luminosity of H1 and ZEUS is about 0.5~fb$^{-1}$.

\section{Supersymmetry}
Supersymmetry~\cite{susy} (SUSY) is by far the most popular theory that extends the Standard Model
since it provides a natural solutions to the hierarchy
problem~\cite{hierarchy1,hierarchy2,hierarchy3} and the unification of gauge forces~\cite{unification} 
and an excellent candidate for the cold dark matter~\cite{dm1,dm2} in the Universe. 

SUSY predicts a partner for each SM particle that carries the same quantum numbers but differs
by half a unit of spin. The lightest SUSY particle (LSP) could be stable if a quantum number called
$R$-parity ($R_p$) is conserved. The primary candidate is currently the lightest neutralino and this is
also an excellent candidate for the Cold Dark Matter. Alternative models suggest the LSP to be 
the gravitino and these are also discussed here. SUSY particles
then typically cascade decay through SM particles to the LSP which escapes detection (if $R_p$ is
conserved).

\subsection{Generic Squarks and Gluinos}
\begin{figure}[h]
\includegraphics[width=0.55\textwidth]{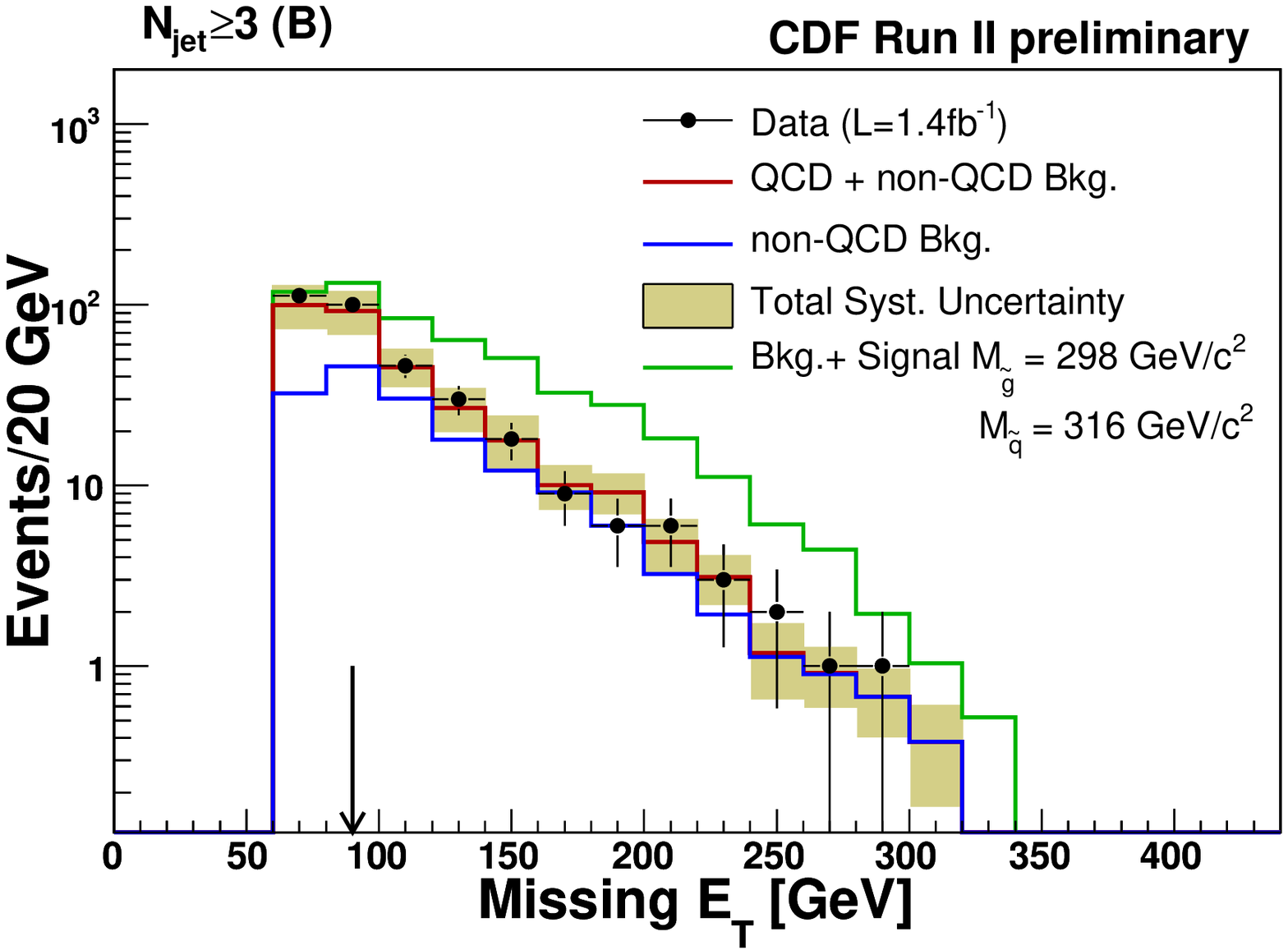}
\includegraphics[width=0.4\textwidth]{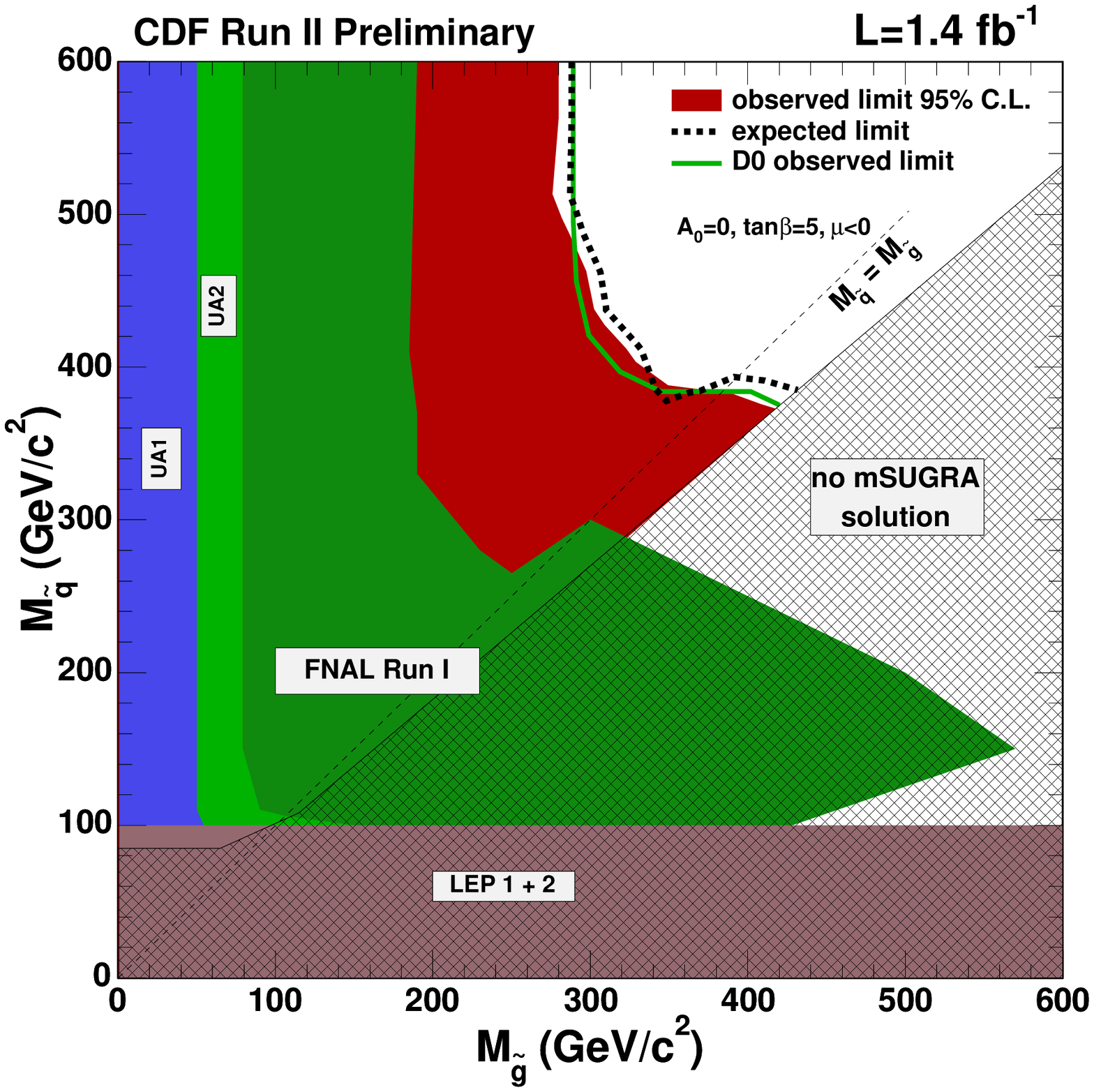}
\caption{\label{fig:susymet} Left: The $\met$ distribution for the 3-jet
final state in the squark/gluino search of CDF. 
Right: The exclusion limits at 95\% CL in the plane
of gluino and squark mass: shown are the limits from UA1, UA2, LEP, 
Run I of the Tevatron and Run II CDF and D0.}
\end{figure}

At the Tevatron one of the primary targets of SUSY searches are the 
colored particles, squarks and gluinos, that are rather copiously produced 
at hadron colliders. In this search the signature is multi-jet production
(from the quarks from the squark or gluino decay) and large missing transverse
energy (due to the neutralinos). This search has been made by both the 
CDF and D0 collaborations: in both cases the experimental collaborations
have searched in the 2-jet+$\met$, the 3-jet+$\met$ and the 4-jet+$\met$ signatures
which are more or less dominant depending on the relationship between the 
gluino and squark mass. If $m(\tilde{g}) \ll m(\tilde{q})$ the dominant process is gluino-pair
production resulting in a 4-jet final state. If $m(\tilde{g}) \gg m(\tilde{q})$ the process
is dominated by squark pair production, yielding a final state of 2 jets 
and $\met$. As an example the $\met$ distribution for the 3-jet final state which is dominant
if $m(\tilde{g}) \approx m(\tilde{q})$ is shown in Fig.~\ref{fig:susymet} for CDF. The dominant
SM backgrounds arise from $W$+jets, $Z$+jets and $t\bar{t}$ production, and a difficult
further background arises from QCD multi-jet
production where $\met$ arises due to severe mismeasurements of the jets.

Both CDF and D0 have performed this search~\cite{cdfd0sq} using $1.4$ and 
$1.0$~fb$^{-1}$, respectively, and neither experiment found any excess 
indicating the presence of squarks or gluinos. Thus both experiments set
lower limits on the masses of these particles as seen in 
Fig.~\ref{fig:susymet}: the gluino is constrained to be heavier
than $290$~GeV/$c^2$ at 95\% CL. 

\subsection{Sbottom and Stop Quarks}
For the above search interpretation the SUSY partners of the 3rd generation 
quarks were not considered since they would introduce a undesired model dependence
and targeted searches are designed for these.
The stop quark is particularly critical in supersymmetry as it is the one
that cancels the large radiative corrections to the Higgs mass from the 
top quark. Due to the large mass of the SM top quark the stop quark mass
splitting is large, and one of the two mass eigenstates is
expected to be the lightest of all squarks:

$$m^2_{\st_{1,2}} = \frac{1}{2} (m^2_{\st_{L}} + m^2_{\st_{R}} \mp 
\frac{1}{2} \sqrt{(m^2_{\st_{L}} - m^2_{\st_{R}})^2 + 4 m_t^2(A_t-\mu\tan\beta)^2}$$

\begin{figure}[h]
\includegraphics[width=0.5\textwidth]{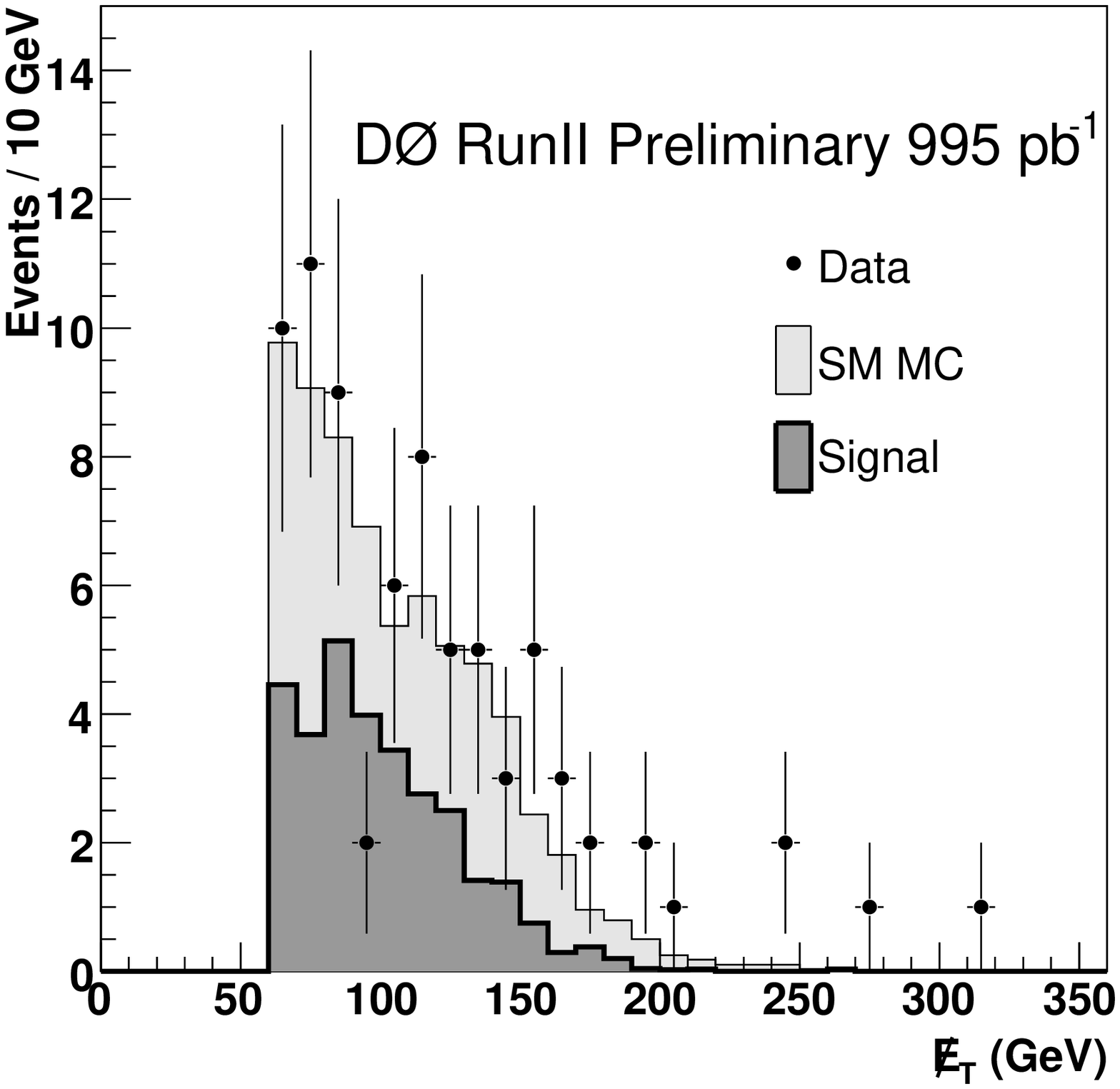}
\includegraphics[width=0.5\textwidth]{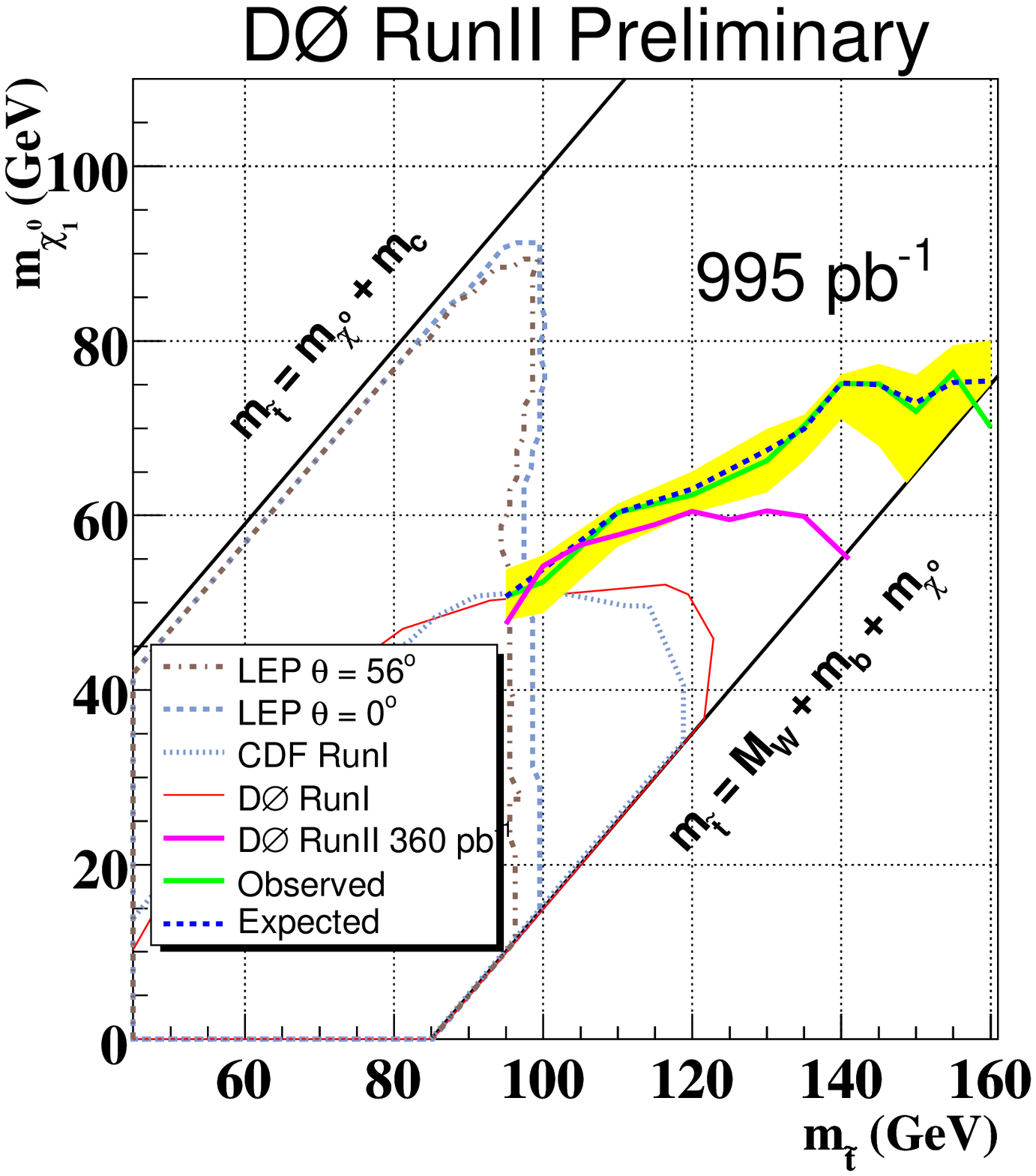}
\caption{\label{fig:stop} Left: The $\met$ distribution for 
$pp \to \st \bar{\st}+X  \to cc\chin\chin +X$ search of D0. Right:
$m(\st)$ vs $m(\chin)$ is shown. Displayed are the kinematic
boundaries of this search and the exclusion limits from CDF, D0 and the LEP
experiments. the yellow band in the D0 analysis shows the uncertainty due to theoretical uncertainties
on the cross section predictions.}
\end{figure}

The Tevatron is most sensitive to the direct pair production, 
$p\bar{p} \to \st \bar{\st} +X$ and many different decay
topologies are possible depending on the masses of the $\st$ and the 
$\chip$ and $\chin$. If $m(\st)>m(\chin)+m(t)$ the decay will proceed 
through $\st \to t\chin$ but the Tevatron experiments are not yet 
sensitive to this case since the cross section in this mass range are too 
low. If $m(\chip)+m(b)<m(\st)<m(\chin)+m(t)$ the decay proceeds
via $\st \to b\chip \to b \ell \nu \chin$ and if this decay is also 
forbidden kinematically it goes via $\st \to c \chin$. These last two decay 
modes have both been searched for (see Ref.~\cite{stopll,stopcdf,stopd0}
and the most recent search~\cite{biscarat} in the $\st \to c \chin$ mode
by the D0 collaboration is shown here. Fig.~\ref{fig:stop} shows the $\met$ distribution
for events with two charm-tagged jets and a variety of other cuts that are designed to
reduce the instrumental background from multi-jet production and backgrounds 
from $W/Z$+jets and $t\bar{t}$ production (for details see. Ref.~\cite{stopd0}). 
The data agree with the SM prediction and limits are placed in the plane of $m(\st)$ versus $m(\chin)$.
For e.g. $m(\chin)=63$~GeV/$c^2$ the limit on the stop mass is $>149$~GeV/$c^2$.

\subsection{Charginos and Neutralinos}
Another class of interesting searches are the searches for direct production
of charginos and neutralinos. Within models where the LSP
is the lightest neutralino a promising search strategy
is the production of three leptons and large $\met$. In gauge mediated SUSY breaking 
(GMSB)~\cite{gmsb1,gmsb2} 
type models where the LSP is the gravitino the signature of $\gamma\gamma+\met$
has been explored.

\begin{figure}[h]
\includegraphics[width=0.5\textwidth]{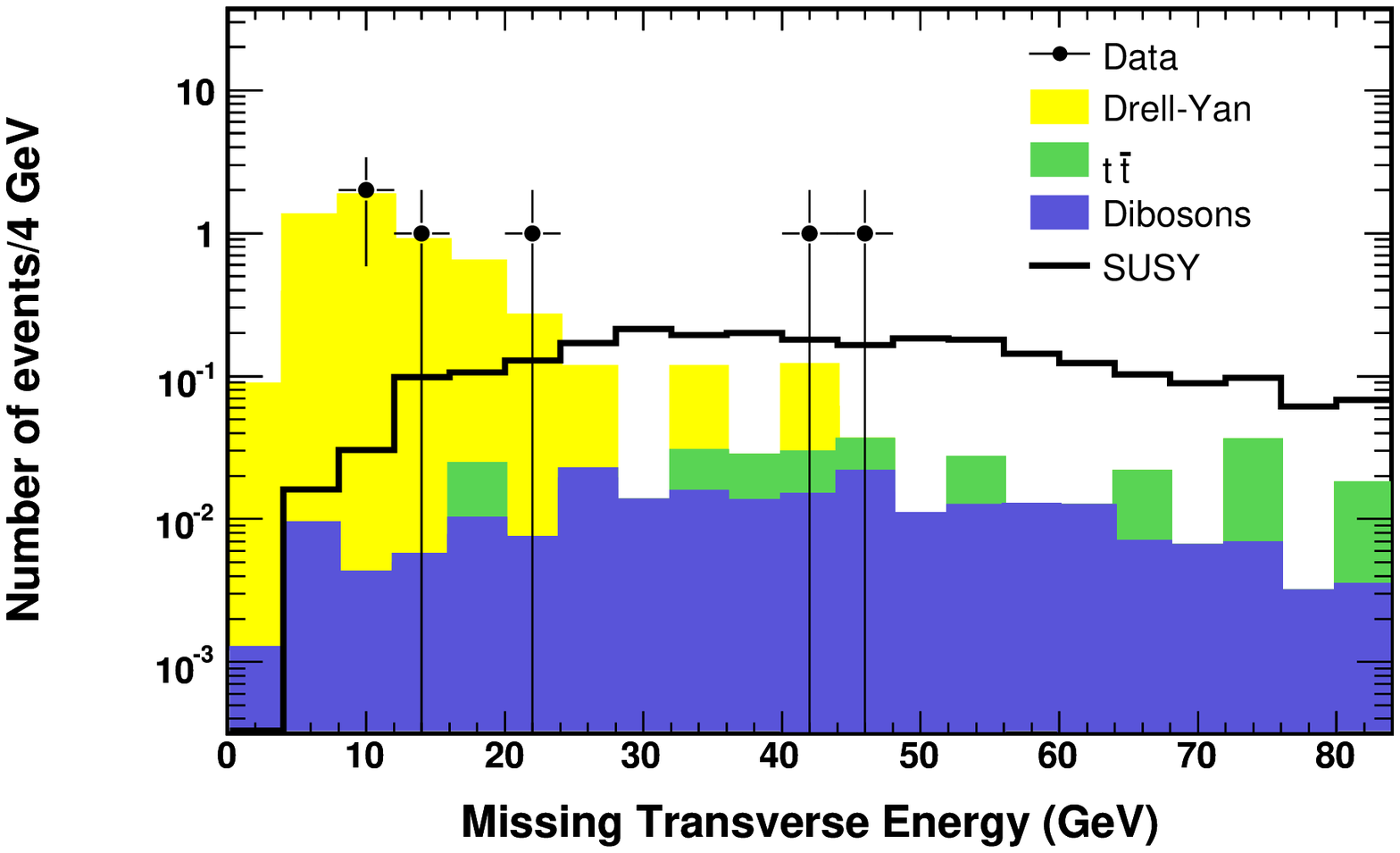}
\includegraphics[width=0.5\textwidth]{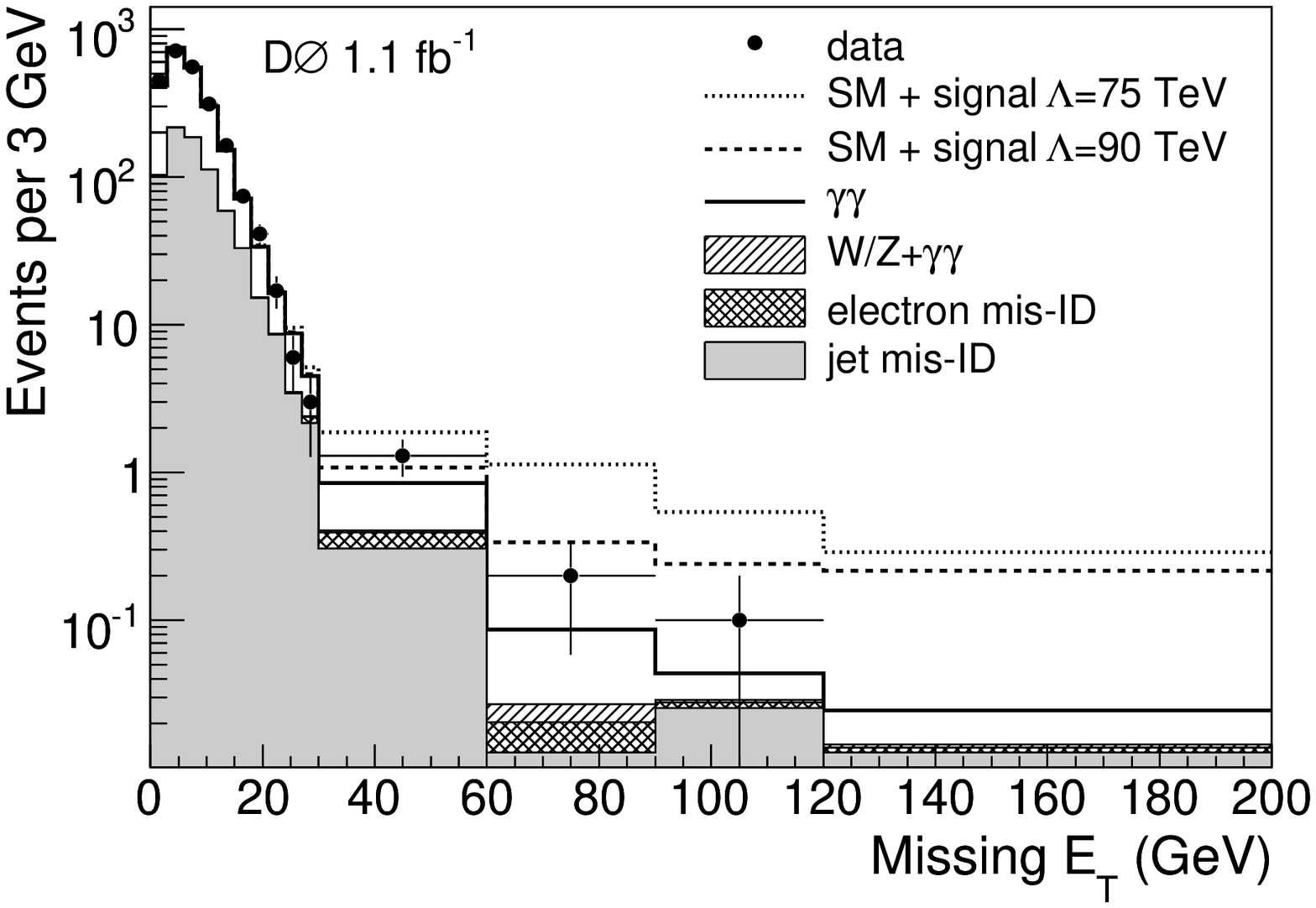}
\caption{\label{fig:chichi} Left: The $\met$ distribution for 
the CDF trilepton search $pbar{p} \to \tilde{\chi_2^0}\chip +X \to 3\ell \chin\chin +X$. 
This example show the search for two electrons and 
an isolated charged track. Right: The $\met$ distribution for diphoton
events as observed by D0. In both figures the data (markers) and the 
backgrounds contributions (histograms) are shown. Also shown are example signal distributions.}
\end{figure}

The trilepton search is unfortunately relatively model-dependent
since the acceptance largely depends on the lepton flavor (in particular 
on the $\tau$ admixture) and the kinematic properties of the leptons, 
in particular their transverse momentum. The searches are divided into 
many sub-signatures to obtain maximum sensitivity. An example of the $\met$
distribution in one of the search signatures is shown in Fig.~\ref{fig:chichi}.
Neither CDF nor D0 observes any significant excess above the SM 
expectation~\cite{d0tril,cdftril} and limits
are placed on the cross section times branching ratio versus chargino mass.
The limits are about 0.1-0.2~pb and constrain the chargino mass in specific
model up to $144$~GeV/$c^2$.

The class of GMSB models where the next-to-lightest SUSY particle (NLSP) 
is the $\chin$ and that decays to $\tilde{G}+\gamma$ is experimentally
relatively easy to observe as there are no intrinsic SM backgrounds 
that give the same signature. D0 has searched for this signature using
1~fb$^{-1}$ of data and find 4 events with $\met>60$~GeV compared
to $1.5 \pm 0.4$ expected from backgrounds. The full $\met$ spectrum is 
shown in Fig.~\ref{fig:chichi}. A lower limit on the 
chargino mass of $231$~GeV is placed. 

\subsection{Long-lived Particles}
Recently a special interest has developed in searches for long-lived 
particles, i.e. particles that live for a while before they decay (see Ref~\cite{longlived} for a
review). These lead to distinct experimental signatures depending on their lifetime and their
exact properties:
\begin{itemize}
\item they can traverse the full detector without stopping. The experimental
signature is a slow charged particle that is tracked through the
tracking detector and triggered in the muon system.
Depending on whether they are strongly
or weakly produced they can undergo charge transformations within the 
detector. For strongly interacting particles this transformation
leads to an effective suppression by a factor 4 per particle. This signature is e.g. produced
by gluinos that can be very long-lived in split-SUSY~\cite{splitsusy} or 
stable $\tilde{\tau}$-leptons in GMSB type models~\cite{gmsbstau}.
\item they can get stuck in the calorimeter and decay there at a later time.
The experimental signature here is a large energy deposition out of time
with any collision. This signature is also predicted by gluinos in split-SUSY and
called ``stopped gluinos''.
\item they can travel for a bit but still decay inside the tracking volume 
before the calorimeter if the lifetime is of the order of about $10$~ns. 
This is predicted e.g. in GMSB models where the lightest neutralino could have a lifetime
and thus travel a significant amount of time before it days to a photon and a gravitino.
\end{itemize}

\begin{figure}[h]
\includegraphics[width=0.33\textwidth]{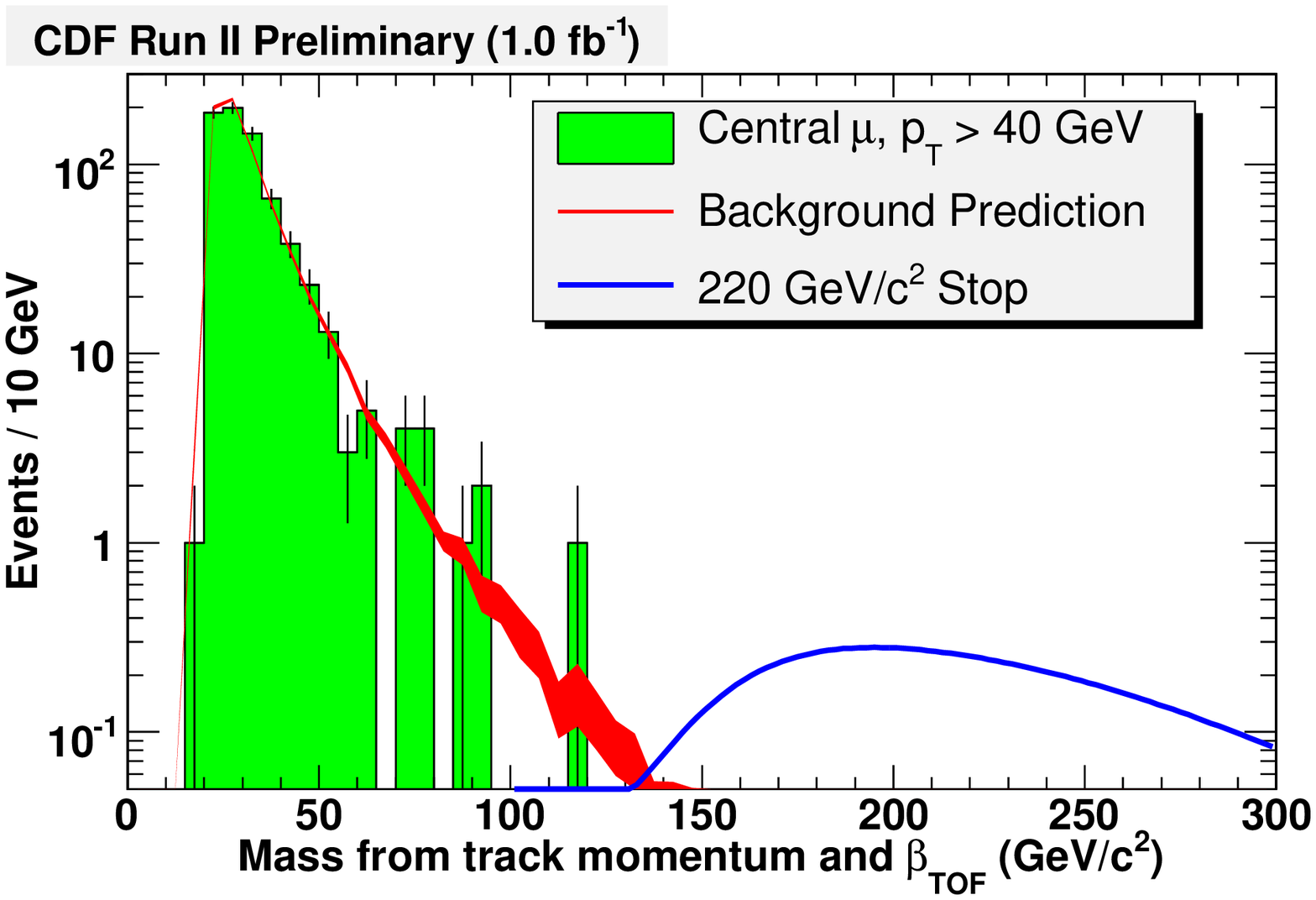}
\includegraphics[width=0.33\textwidth]{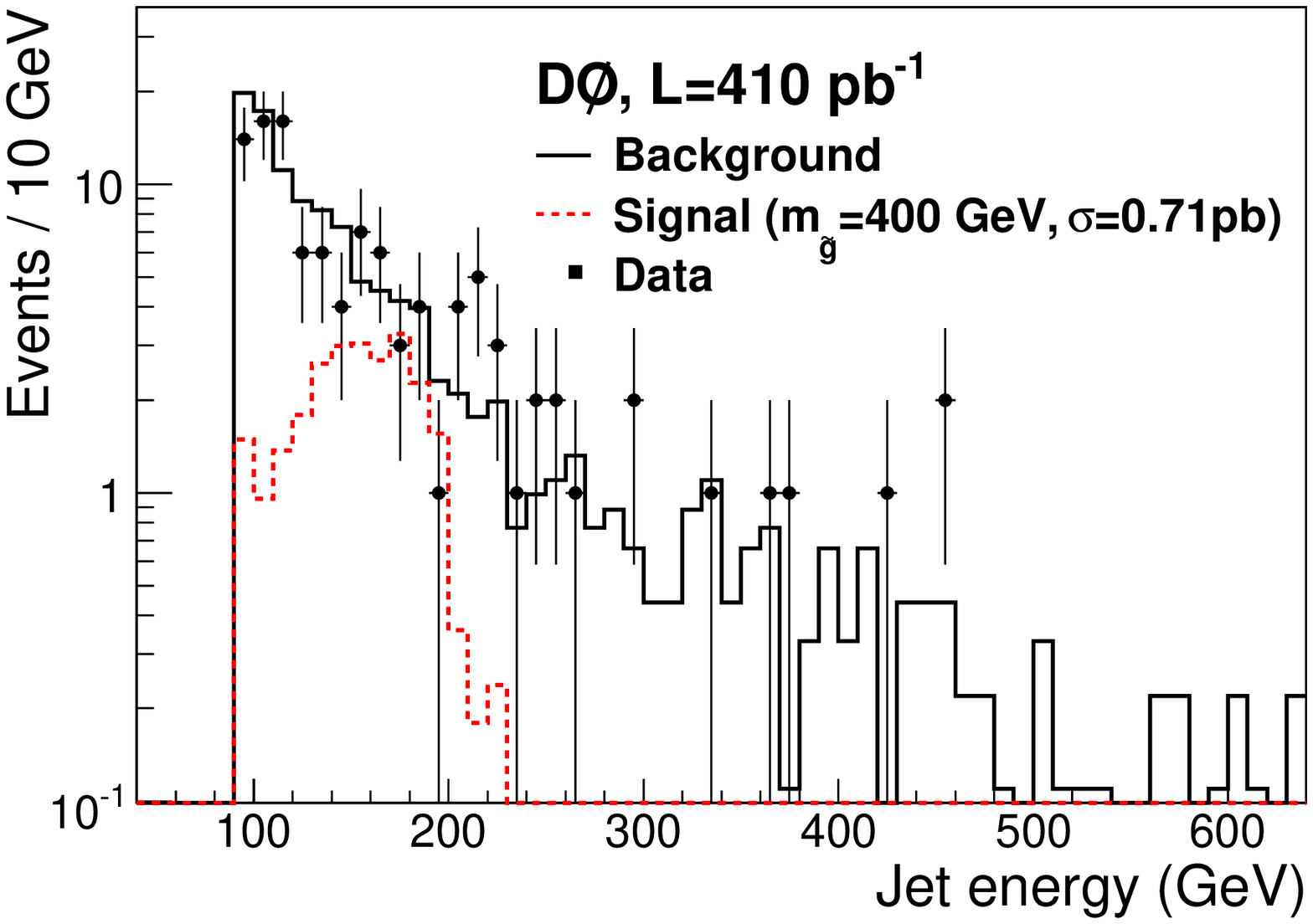}
\includegraphics[width=0.33\textwidth]{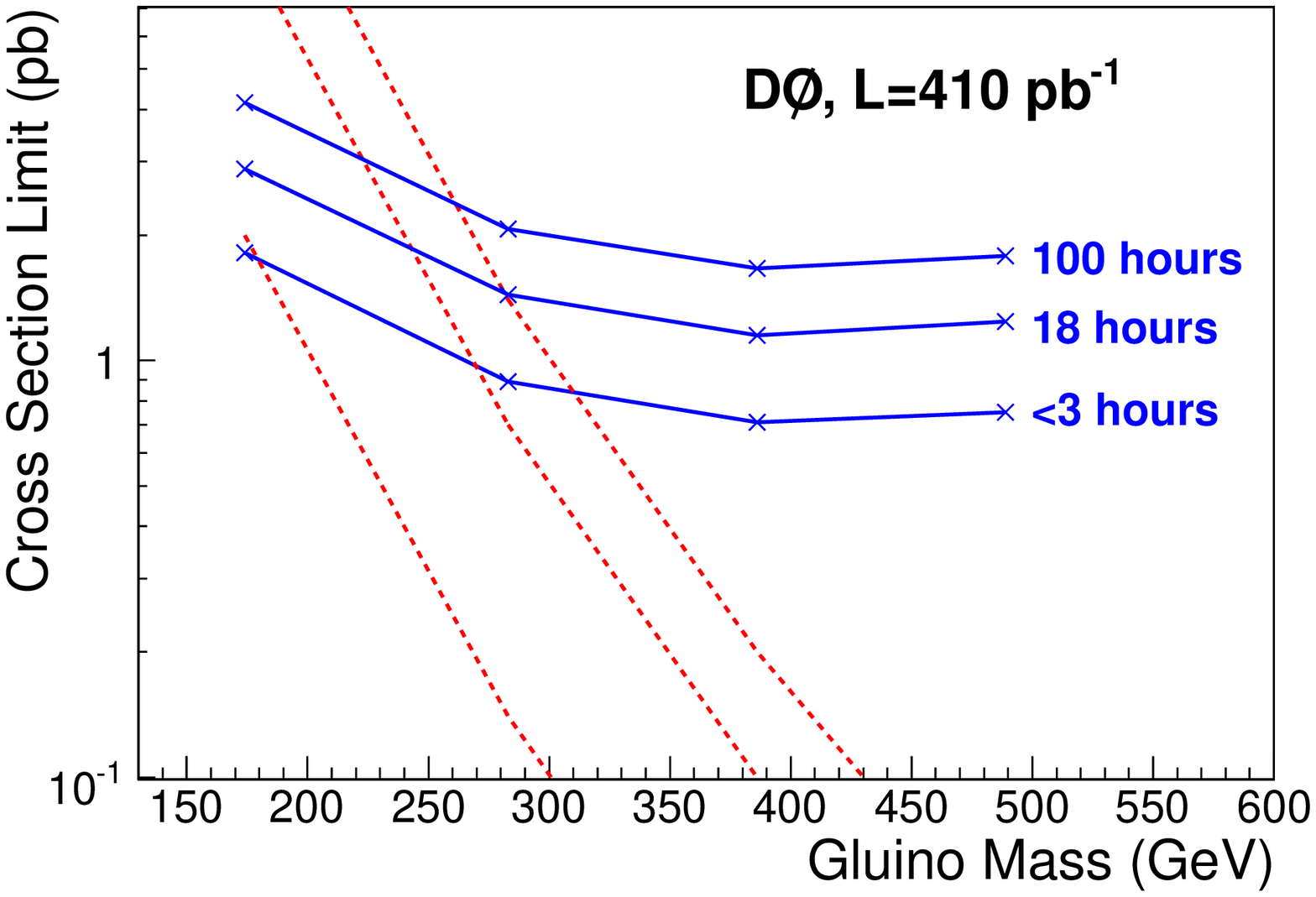}
\caption{\label{fig:champ} Left: Mass distribution for the CDF search for a long-lived charged 
particle traversing the detector. Middle: Jet energy distribution for D0's search for a long-lived
particle that later decays in the detector out of time with any collision. Right: Cross
section time branching ratio for stopping gluinos. Theoretical predictions (red dashed lines) 
and experimental upper limits (blue solid lines) are shown for different assumptions.}
\end{figure}

CDF and D0 searched for slow particles traversing the full detector, and both 
trigger the particle by a signal in the muon system. CDF
uses the time-of-flight detector just outside the tracking system at a radius
of about 1m to measure the arrival time with a resolution of 100~ps
and reconstructs the mass by $m=p\sqrt{1/\beta^2-1}$. D0 uses the timing
from the muon system itself. For weakly interacting particles CDF observes
no evidence or such particles and places
cross section limits for particles with $p_T>40$~GeV, 
$|\eta|<1$ and $0.4<\beta<0.9$ of 10 fb, for strongly interacting
particles the corresponding limit is 48~fb~\cite{champ}. The mass distribution
is shown in Fig.~\ref{fig:champ}. As an example this constrains 
a stable stop quark to have a mass greater than 250~GeV/$c^2$. D0
has carried out a similar search~\cite{champ}.

D0 have also searched for so-called ``stopped gluinos''
by looking for events where no interaction is seen
but there is a large energy deposit in the calorimeter~\cite{stopgluino}. 
The main backgrounds are cosmic ray and beam halo muons that shower in the calorimeter.
Fig.~\ref{fig:champ} shows the data compared to the SM expectation 
and a hypothetical signal. Since there is no sign of any deviation from the SM
limits are placed depending on the assumed lifetime. These are shown
in Fig.~\ref{fig:champ} and compared to the predicted cross
section. The analysis probes gluino masses between 175 and 320 GeV depending
on the details of the model.

CDF has also searched for photons that arrive out of time, so-called ``delayed photons''~\cite{delpho} 
which can stem from a long-lived massive particle. CDF measures the arrival time o the photon
in the electromagnetic calorimeter. The data show no sign of any excess and are used to 
constrain the neutralino mass to be larger than $101$~GeV/$c^2$ for a lifetime $\tau_{\chin}=5$~ns.

\subsection{Higgs boson}

Within the minimal supersymmetric Standard Model (MSSM) 
there are two scalar fields resulting in a total of five physical
Higgs bosons: the scalar h and H, the pseudo-scalar A and the two charged 
states $H^\pm$. At high $\tan\beta$ the $A$ is degenerate with either the
$h$ or the $H$ or with both and additionally the cross section is 
enhanced~\cite{mssmhiggs1,mssmhiggs2,mssmhiggs3} with $\tan^2\beta$:
$$\sigma = 2 \sigma_{SM} \frac{\tan^2\beta}{(1+\Delta_b)^2} 
\frac{9}{[9+(1+\Delta_b)^2]}$$

\begin{figure}[h]
\begin{minipage}{20pc}
\includegraphics[width=20pc]{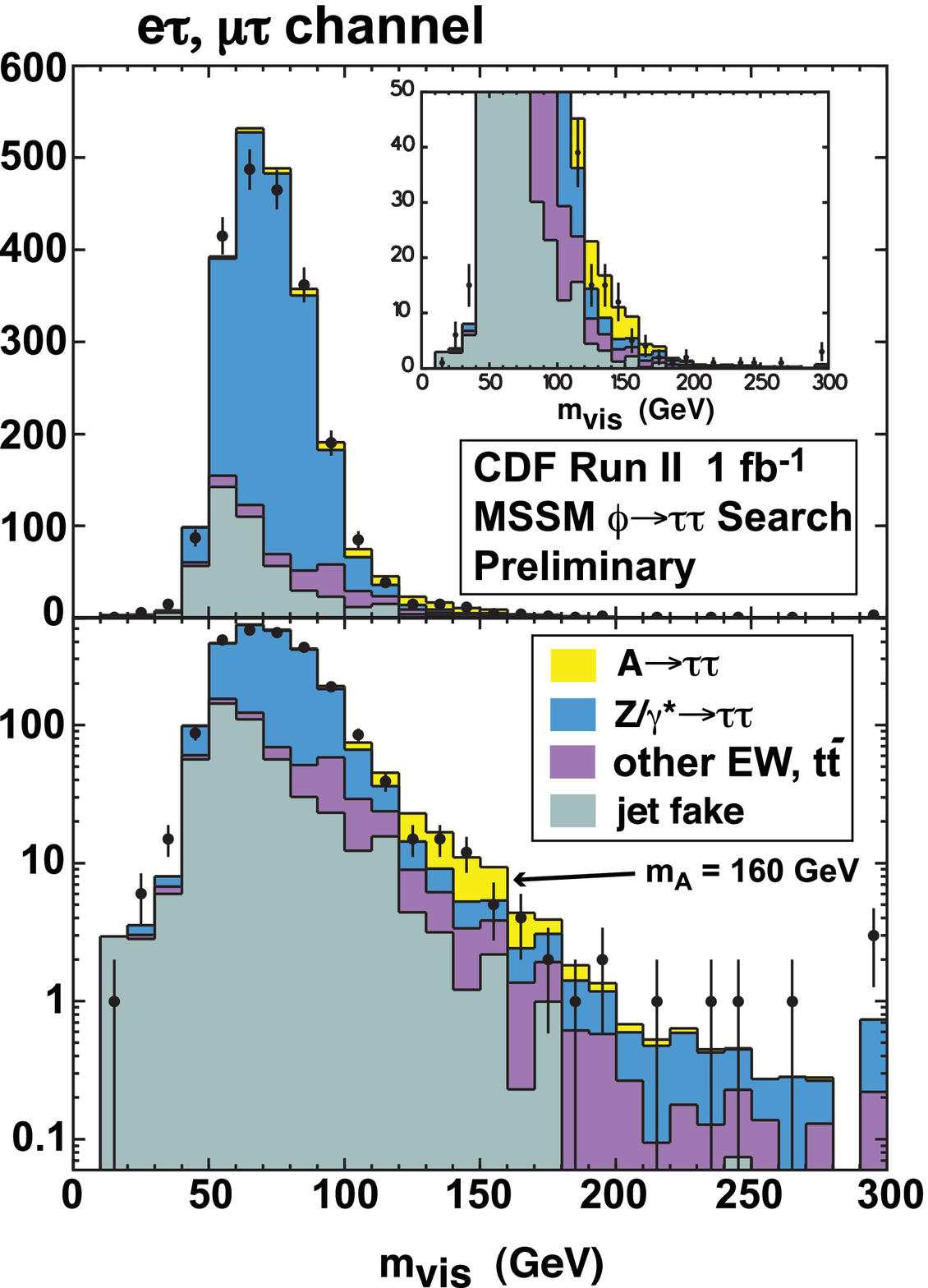}
\caption{\label{fig:mssmhcdf} CDF's $m_{vis}$ distribution for the MSSM Higgs 
search in the di-tau decay mode ($\phi \to \tau^+ \tau^-$ with $\phi=A,h,H$). }
\end{minipage}\hspace{2pc}%
\begin{minipage}{16pc}
\includegraphics[width=16pc]{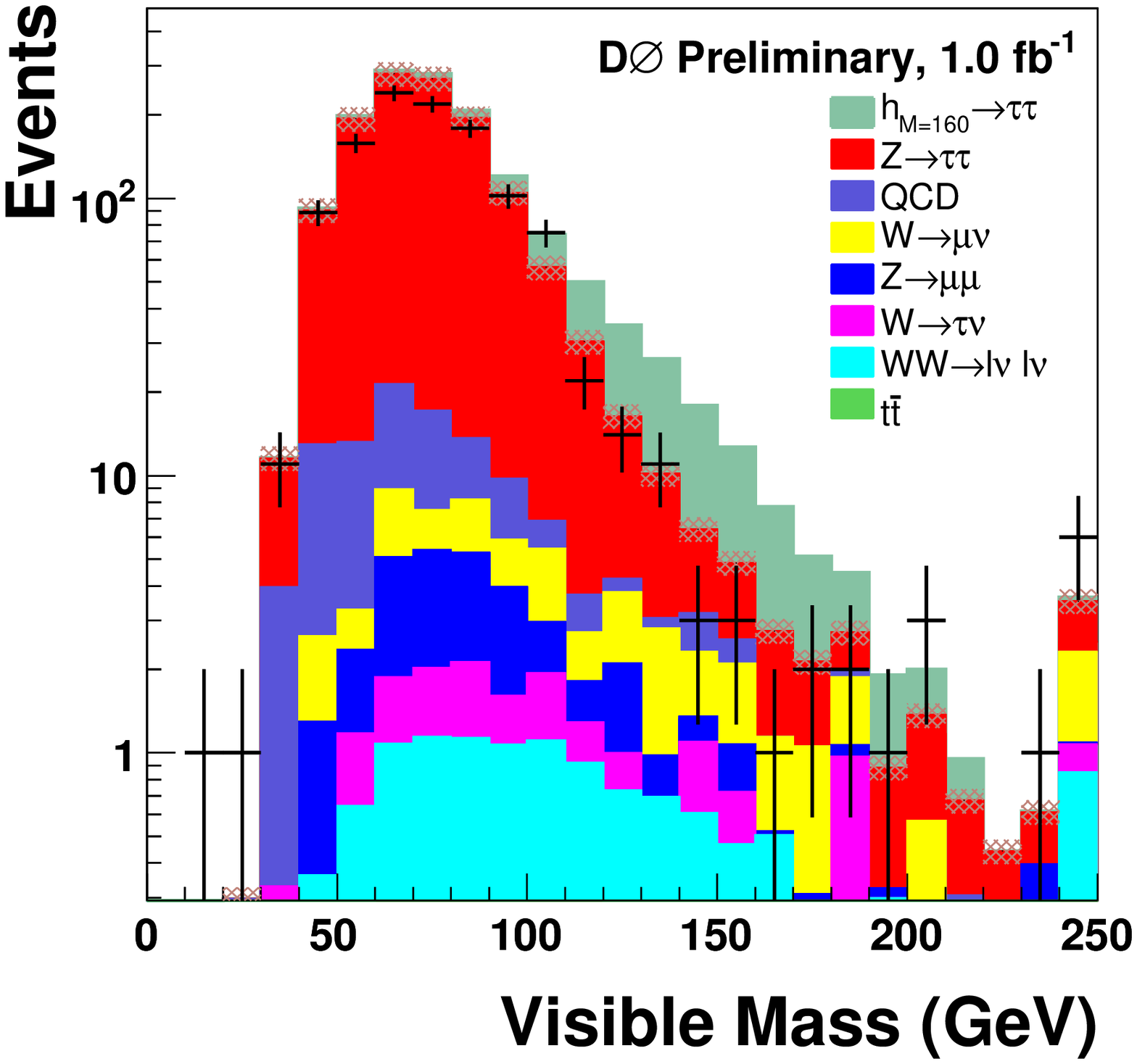}
\includegraphics[width=16pc]{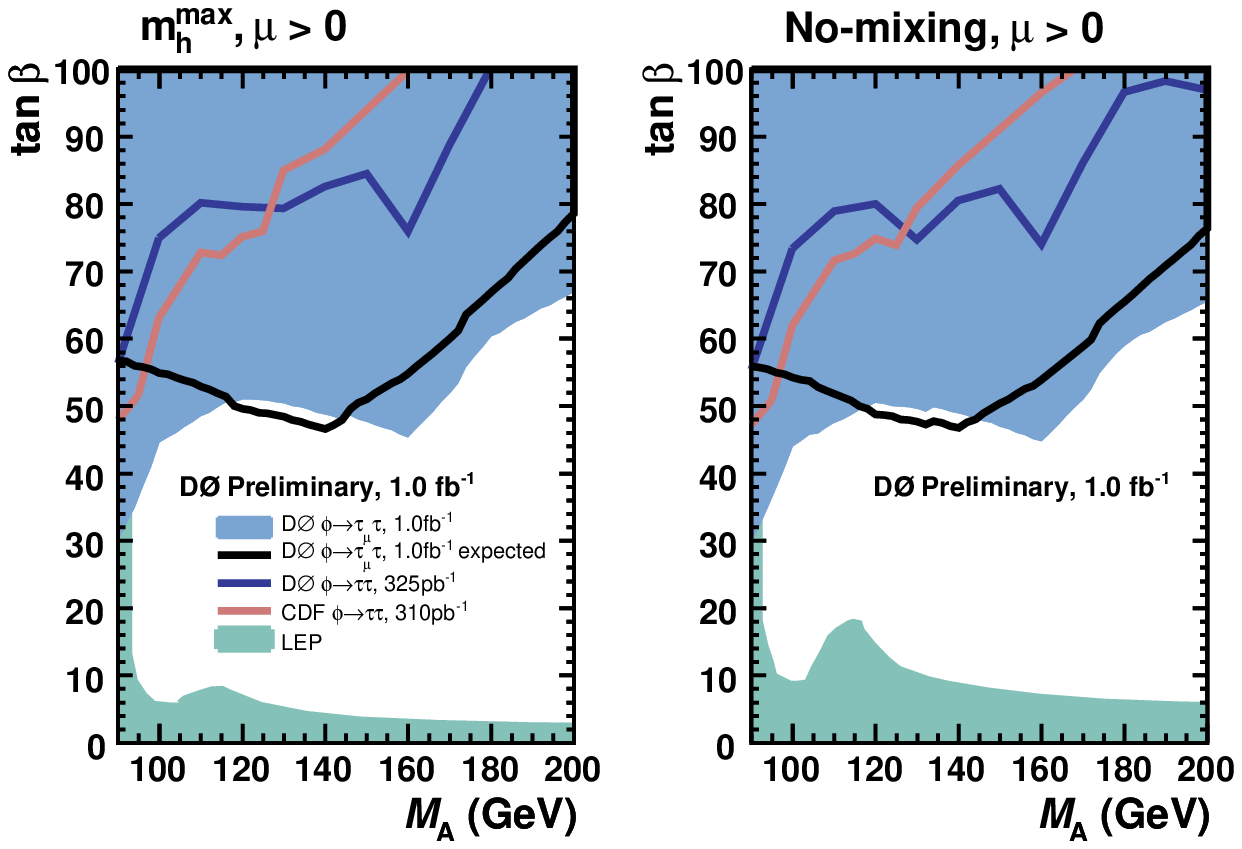}
\caption{\label{fig:mssmhd0} Top: D0's $m_{vis}$ distribution for the MSSM Higgs 
search in the di-tau decay mode ($\phi \to \tau^+ \tau^-$ with $\phi=A,h,H$). 
Bottom: Plane of $\tan\beta$ 
versus $m_A$: the excluded regions is in light blue. Also shown are previous results and the expected 
limit. The limits are shown in a specific scenario called ``no-mixing''~\cite{carena}}
\end{minipage} 
\end{figure}

At high $\tan\beta$ the Higgs boson decays about $90\%$ to $b\bar{b}$ and $10\%$ to $\tau^+\tau^-$. 
The searches
presented here use the $\tau^+\tau^-$ decay modes. The signal is selected by
requiring at least one electron or muon from one of the tau decays 
($\tau_e$, $\tau_\mu$) and then
the other tau can either decay hadronically ($\tau_h$) or also leptonically.
Due to the outgoing neutrinos it is not possible to fully reconstruct
the invariant mass of the two $\tau$'s in this inclusive search. Thus 
a quantity called ``visible'' mass is defined as $m_{vis}=p_T(\tau,1)+p_T(\tau,2)+\met$. This is shown for the CDF and D0 searches in Fig.~\ref{fig:mssmhcdf} and Fig.~\ref{fig:mssmhd0}.
While CDF observed a small excess at $m_A \approx 160$~GeV a slight deficit
is seen by D0. The experiments are sensitive to $\tan\beta \approx 40-60$ 
for $m_A<200$~GeV as seen in Fig.~\ref{fig:mssmhd0}.

Alternative searches probing the MSSM Higgs boson at high $\tan\beta$ are those for multi-$b$-jet 
production which have also been carried out by CDF and D0~\cite{bbb} but also
show no sign of new physics.

Similar parameter space is probed indirectly in rare $B$-decays 
(e.g. $B_s \to \mu^+\mu^-$ by CDF/D0~\cite{aurore} and
$B^+ \to \tau^+ \nu_{\tau}$ by Belle/Babar~\cite{btau}), in $\tau$ decays 
(e.g. $\tau \to \mu\eta$ by Belle/Babar~\cite{tau1,tau2}) and in rare kaon decays (e.g.
in the precision measurement of 
$R_K = \Gamma(K\to \mu\nu)/\Gamma(K\to e\nu)$ by KLOE and NA48~\cite{rarek1,rarek2}.

\section{Beyond Supersymmetry}
Even though Supersymmetry is the most favored model of physics beyond the
SM, there is of course a good chance that it does not represent Nature and
that other particles appear at the TeV scale.

\subsection{Contact Interactions and Compositeness}
A classic way of parameterizing new physics is the introduction of contact
terms that mediate 4-fermion interactions, similar to Fermi's proposal
for explaining $\beta$-decay where the contact term was later explained by the
exchange of a $W^\pm$ boson. The Lagrangian for a contact interaction can most 
generally be written as 

$${\cal L}=\frac{4\pi}{\Lambda^2}\sum_{i,j}\eta_{ij}\bar{q_i \gamma^\mu q_i} \bar{\ell_j \gamma^\mu \ell_j} $$
where $\eta_{ij}$ is a sign factor that is positive (negative) for constructive (destructive) 
interference and $i,j = L,R$ denote left and right-handed helicities of the quarks and leptons.

\begin{figure}[h]
\includegraphics[width=0.55\textwidth]{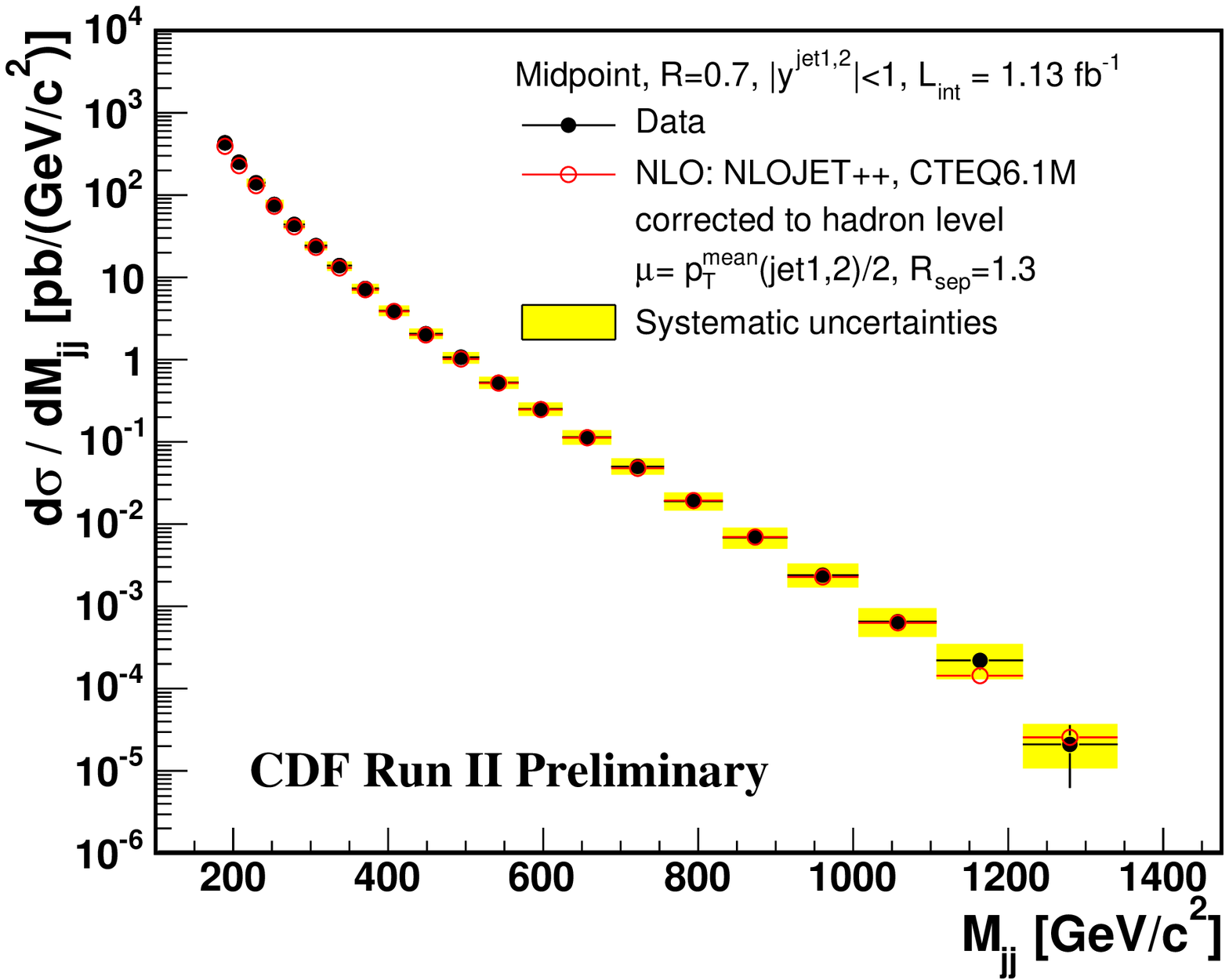}
\includegraphics[width=0.45\textwidth]{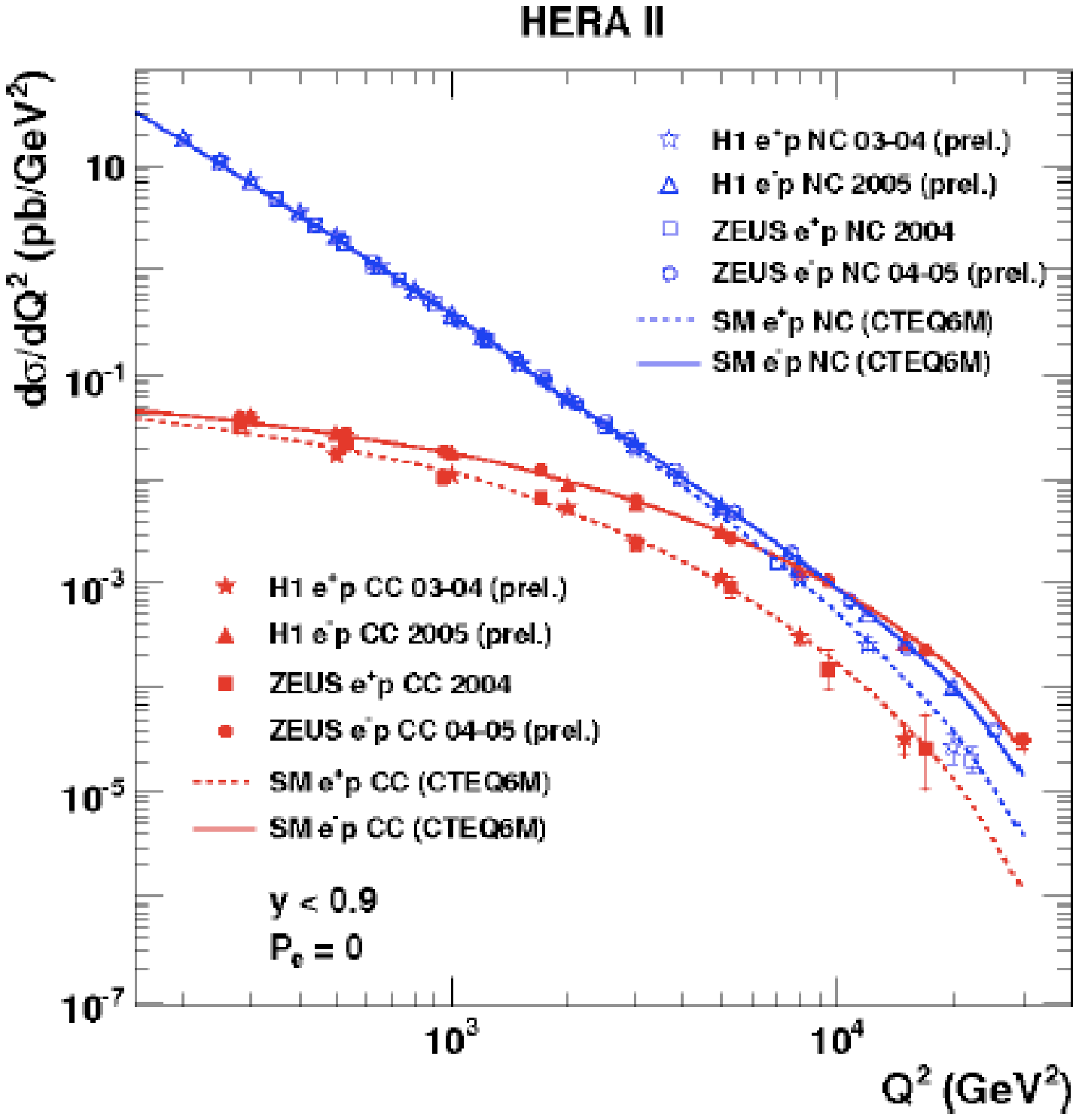}
\caption{\label{fig:inclxs} Left: The dijet cross section versus the dijet invariant mass, $M_{jj}$. 
Right: The neutral and charged current cross section versus $Q^2$ for $e^+p$ and $e^-p$ scattering.}
\end{figure}

Figure~\ref{fig:inclxs} shows inclusive spectra of the dijet cross section
versus the dijet invariant mass at the Tevatron and the Neutral and Charged
Current cross sections at HERA. A beautiful agreement between the data and 
the Standard Model predictions is observed over up to seven orders of 
magnitude. 

These can be used to constrain contact interactions and e.g. for $eeqq$
contact terms some example constraints from D0 and ZEUS are given in 
table~\ref{tab:ci}. 

\begin{table}[h]
\centering
\caption{\label{tab:ci} Limits on $eeqq$ contact interactions terms from the D0
and ZEUS collaborations and limits on $\mu\mu qq$ contact terms from D0. All limits
are given in units of TeV.}
\begin{tabular}{|l|rr|rr|rr|}
\br
Model &    \multicolumn{6}{|c|}{Coupling Structure}\\\hline
  & \multicolumn{2}{|c|}{ZEUS  $eeqq$} & \multicolumn{2}{|c|}{D0  $eqq$} & \multicolumn{2}{|c|}{D0 $\mu\mu qq$}\\
  & $\Lambda^-$ & $\Lambda^+$ & $\Lambda^-$& $\Lambda^+$ & $\Lambda^-$& $\Lambda^+$\\\hline
LL & 4.2 & 4.2  & 6.2 & 3.6 & 7.0 & 4.2\\
LR & 2.0 & 3.6  & 4.8 & 4.5 & 5.1 & 5.3\\
RL & 2.3 & 3.6  & 5.0 & 4.3 & 5.2 & 5.3\\
RR & 4.0 & 3.8  & 5.8 & 3.8 & 6.7 & 4.2\\
\hline
\mr
\end{tabular}
\end{table}

Difermion mass spectra are also sensitive to many specific models, e.g. in models
with large extra dimensions at the TeV scale (ADD~\cite{add}) or models where one of 
the extra dimensions is warped (RS~\cite{rs}) or models predicting new gauge bosons ($Z'$, $W'$).

\begin{figure}[h]
\includegraphics[width=0.31\textwidth]{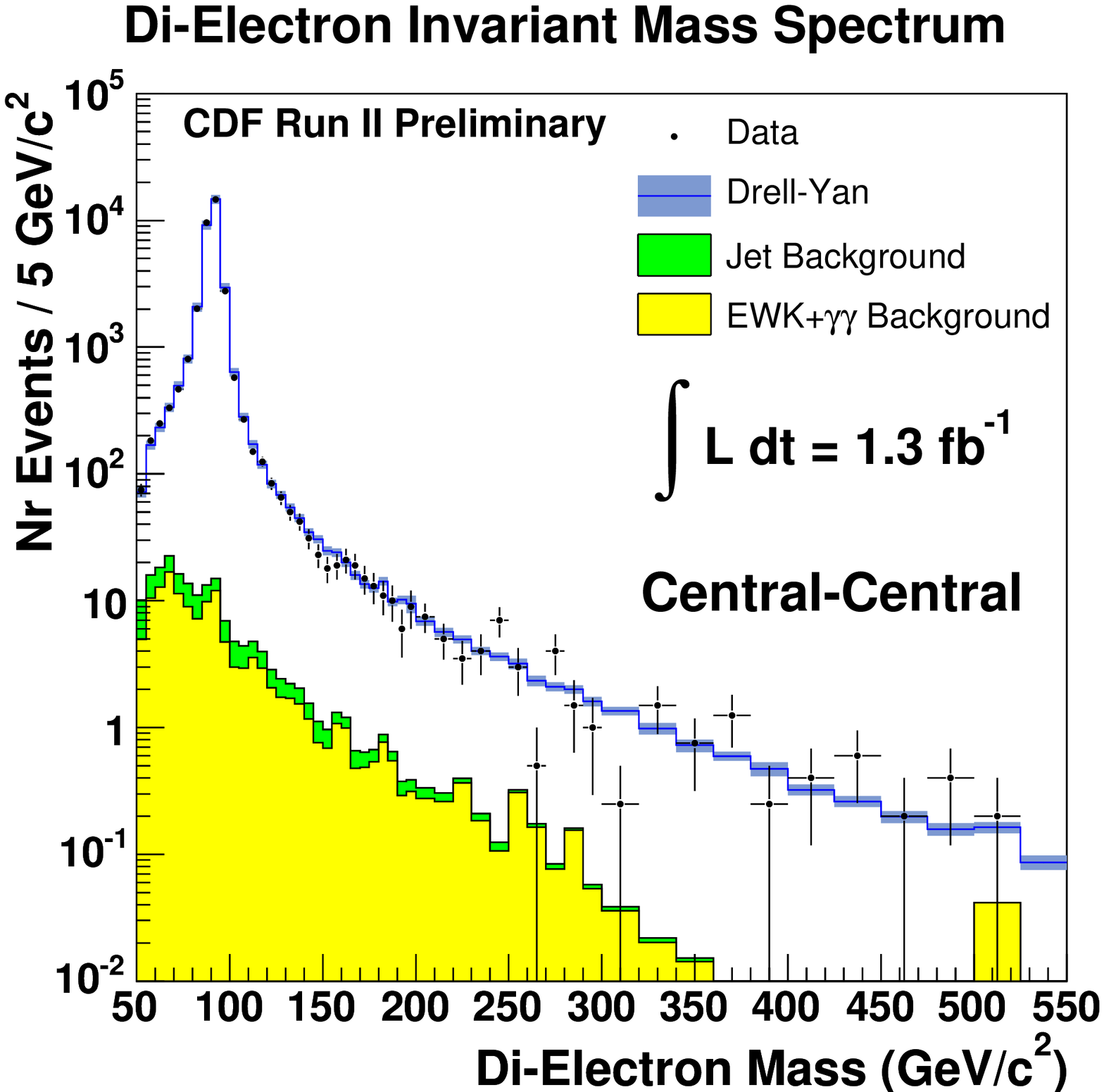}
\includegraphics[width=0.33\textwidth]{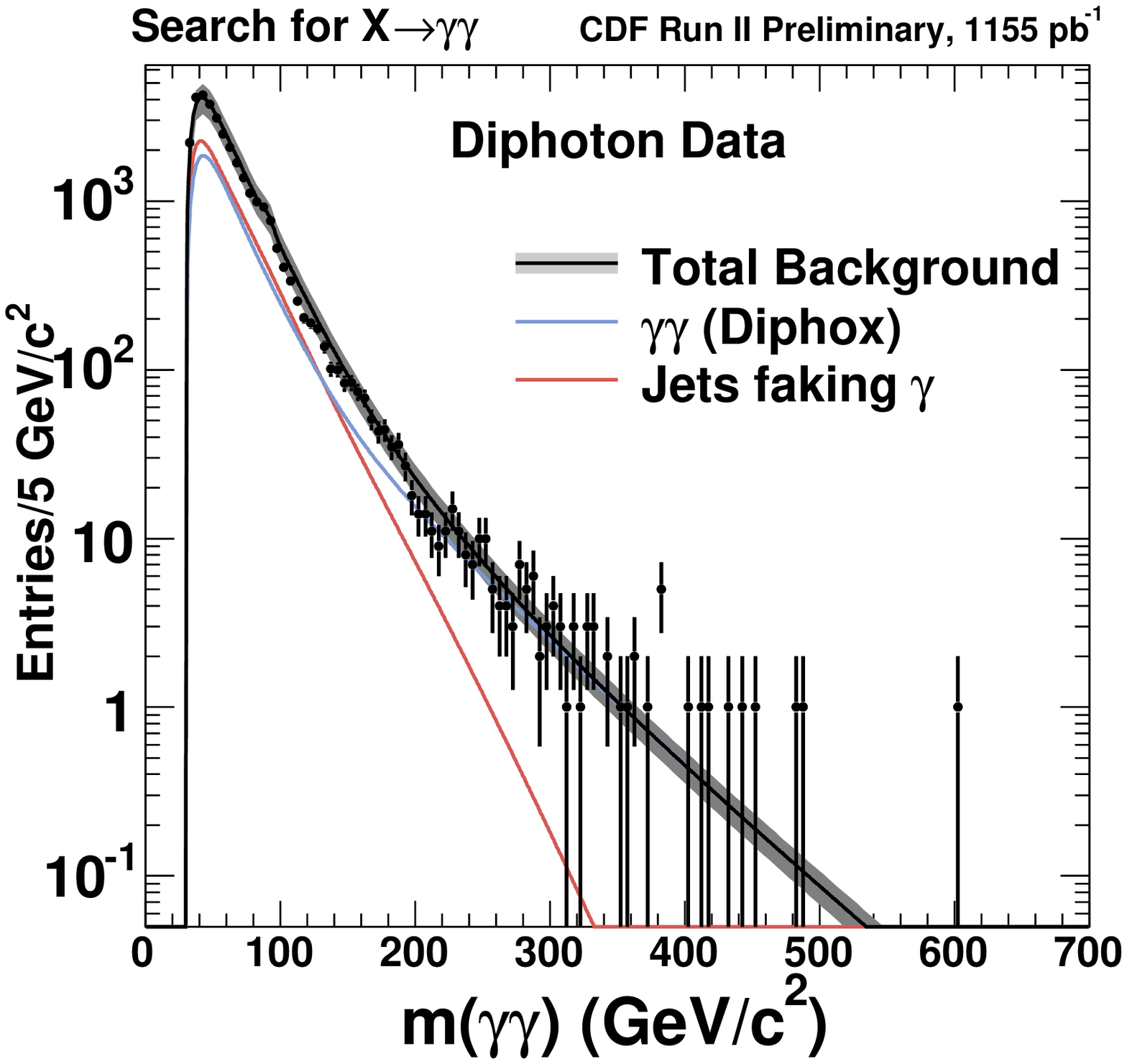}
\includegraphics[width=0.4\textwidth]{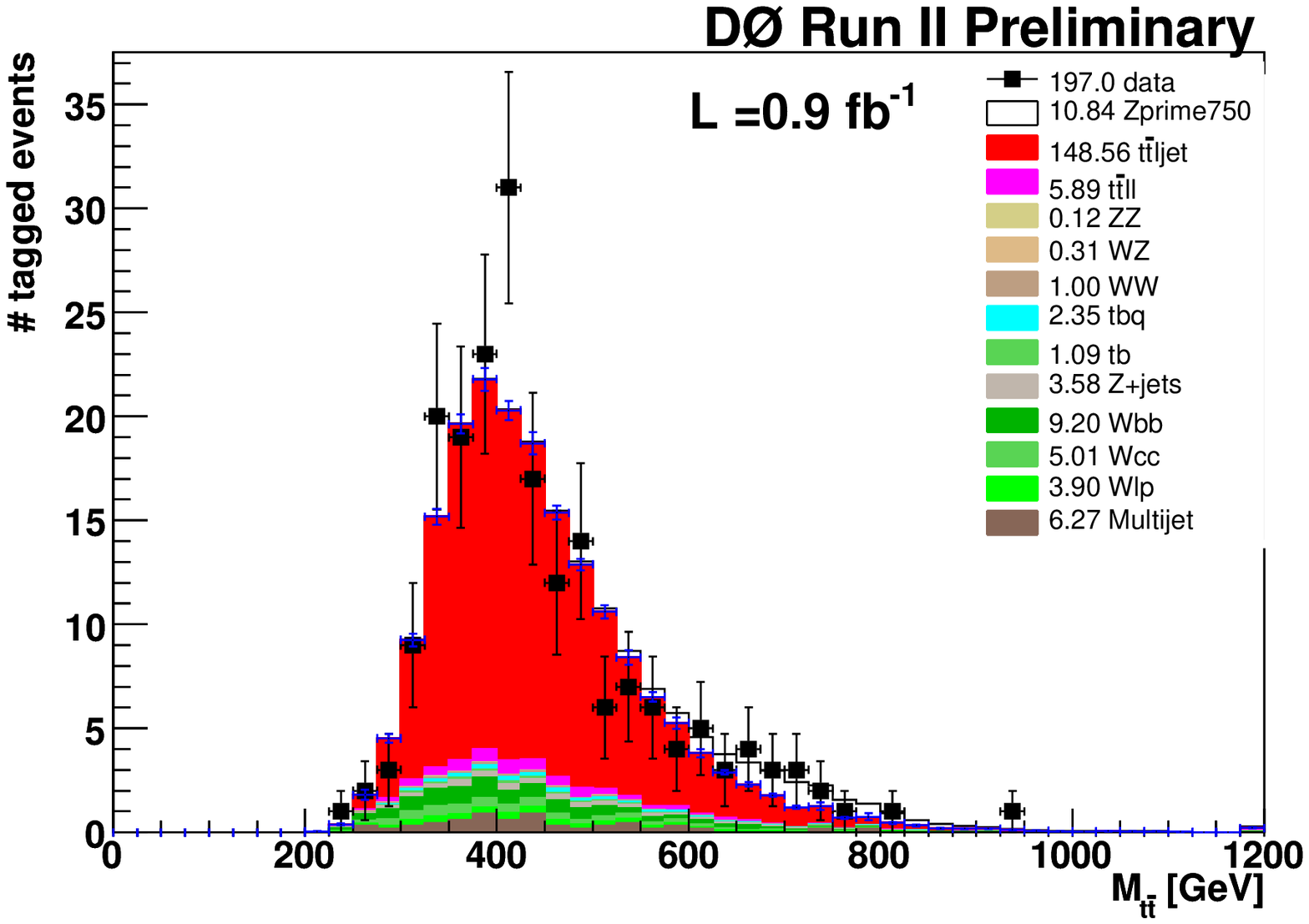}
\caption{\label{fig:zprime} Invariant $ee$ (left) and $\gamma\gamma$ (middle) mass spectrum at CDF, and the
$t\bar{t}$ invariant mass spectrum by D0. In all cases the data (points)
are compared to the SM expectation (histograms).}
\end{figure}

Figure~\ref{fig:zprime} shows the inclusive dielectron, diphoton and the $t\bar{t}$ mass spectra that are
sensitive to new electrically neutral resonances. Using these data a $Z'$ with SM couplings is constrained
to be heavier than $925$~GeV/$c^2$ by CDF~\cite{zprimecdf}, and 
a RS-Graviton $>900$~GeV/$c^2$ for $k/M_{Pl}=0.1$ 
(and $>300$~GeV/$c^2$ for $k/M_{Pl}=0.01$ (see CDF~\cite{zprimecdf} and D0~\cite{d0rs}).
A Z'-boson in topcolor models is constrained to be heavier than $720$~GeV/$c^2$.

\begin{figure}[h]
\includegraphics[width=0.4\textwidth]{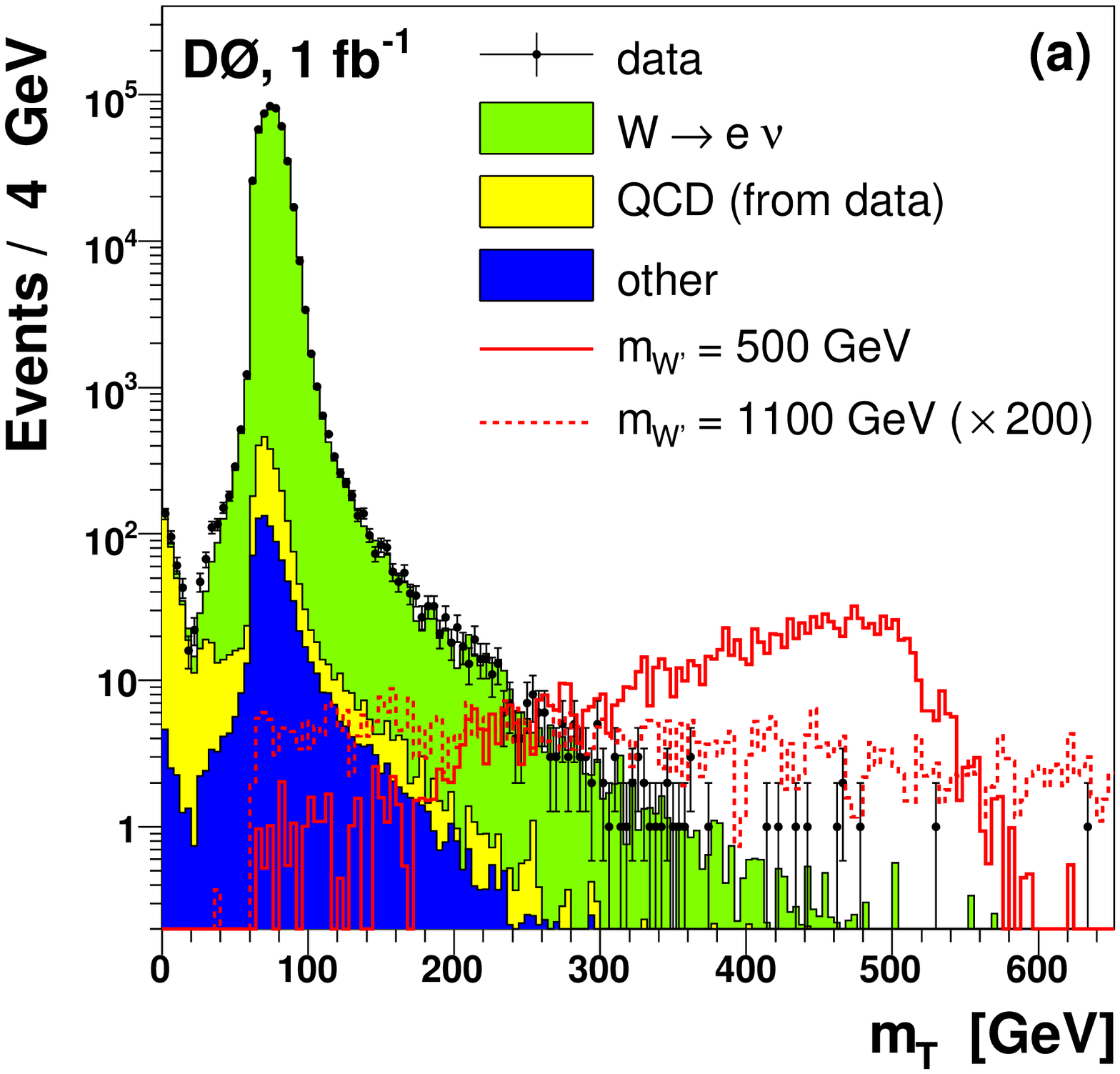}
\includegraphics[width=0.6\textwidth]{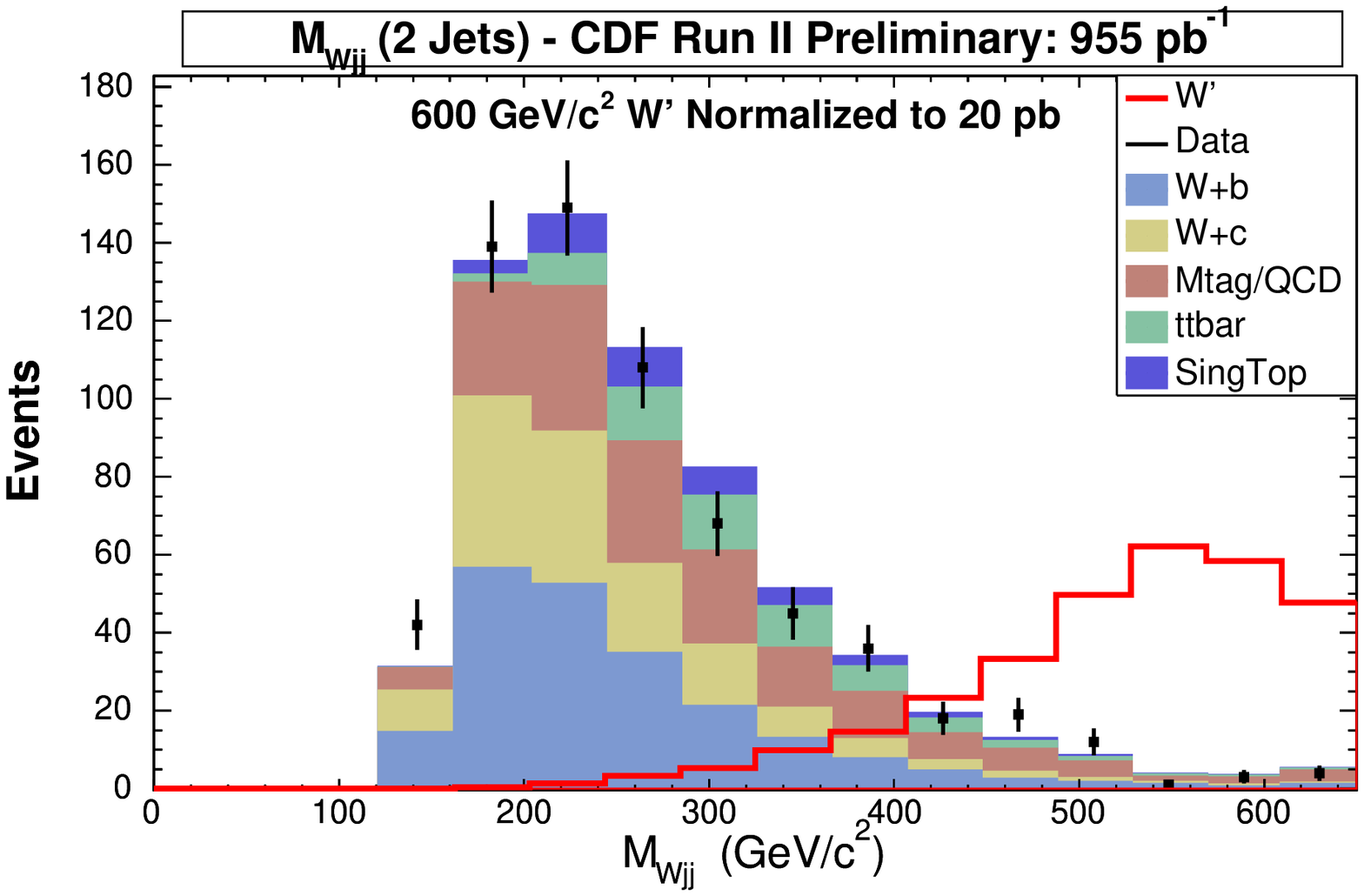}
\caption{\label{fig:wprime} Left: Transverse $e\nu_e$ mass in D0's $W'$ search. 
Right: Invariant mass of $tb$ in CDF's $W'$ search. In both cases the data (points) are compared to the
SM predictions (filled histograms). Example $W'$ signals are also shown as open histograms.}
\end{figure}

Figure~\ref{fig:wprime} shows the transverse mass for $p\bar{p} \to e\nu_e +X$ events 
and the reconstructed mass for $p\bar{p} \to tb +X$ events. Both signatures are sensitive to e.g. $W'$
production and have been used to constrain a $W'$ to have $m(W')>965$~GeV/$c^2$ (using the D0 $e\nu_e$ 
analysis) and $m(W')>760$~GeV/$c^2$ using the CDF $tb$ analysis). Note, that while the $e\nu_e$ only probes
left-handed currents the $tb$ analysis is sensitive to both left- and right-handed currents.

\begin{figure}[h]
\includegraphics[width=0.32\textwidth]{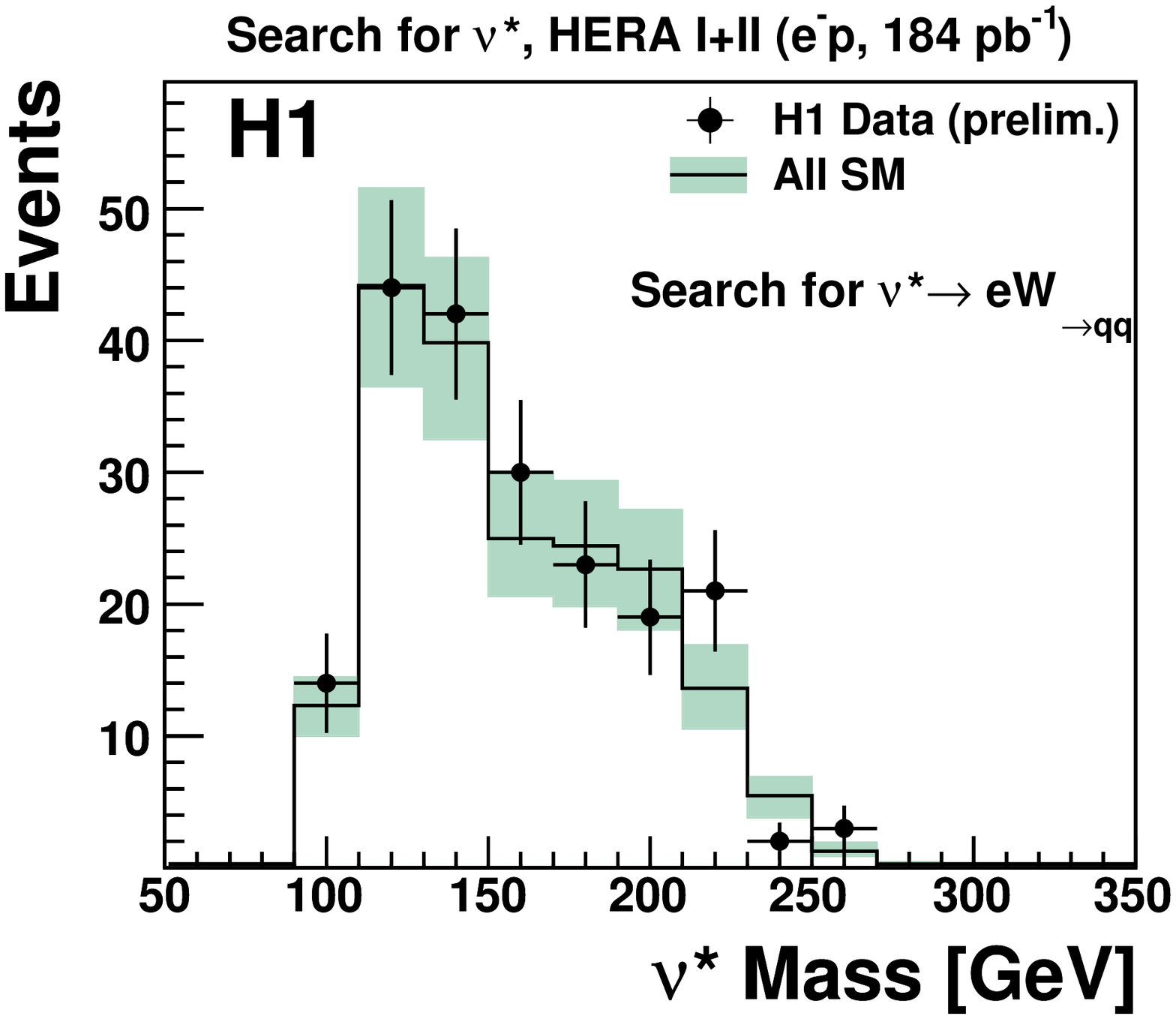}
\includegraphics[width=0.38\textwidth]{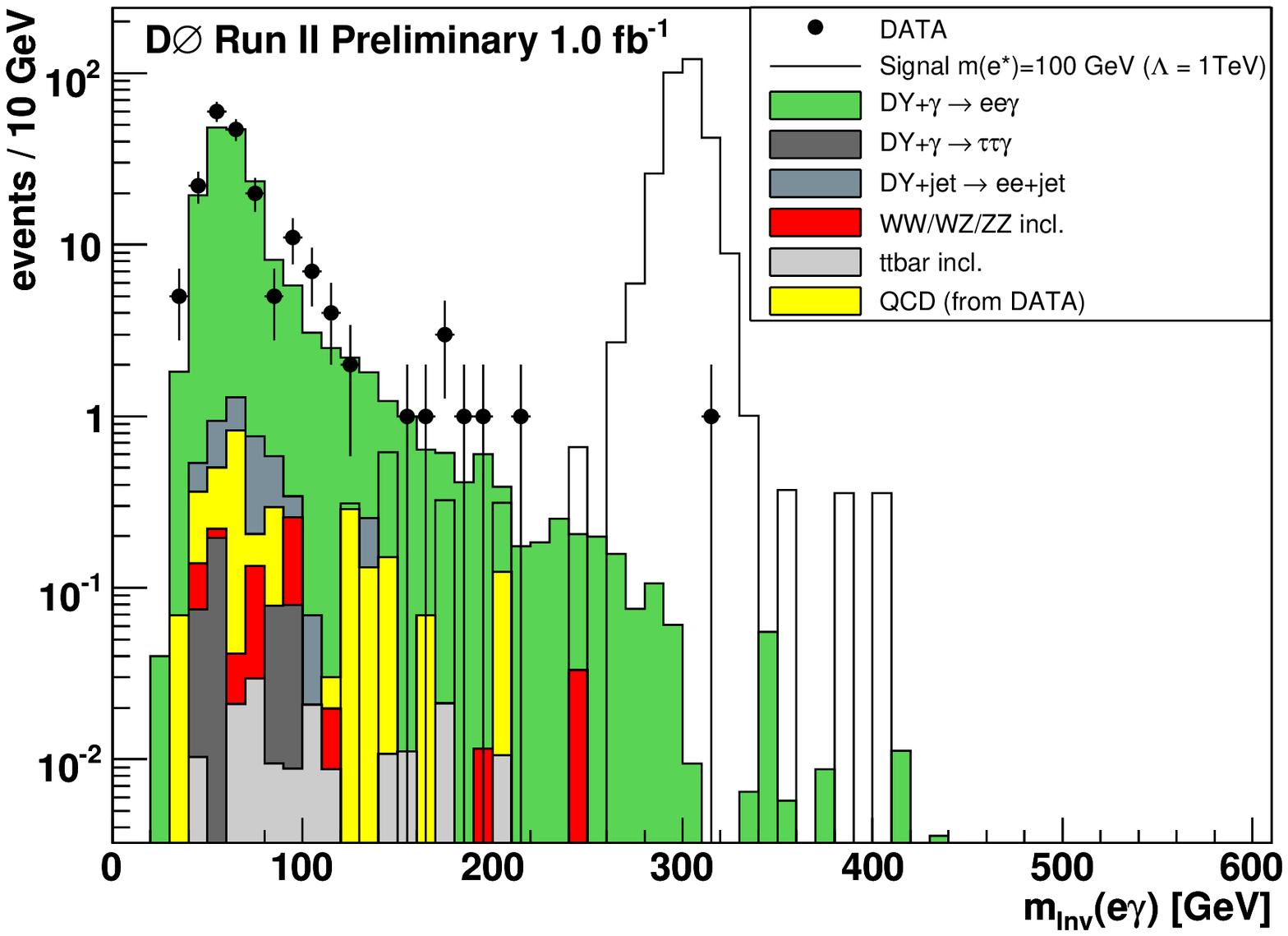}
\includegraphics[width=0.35\textwidth]{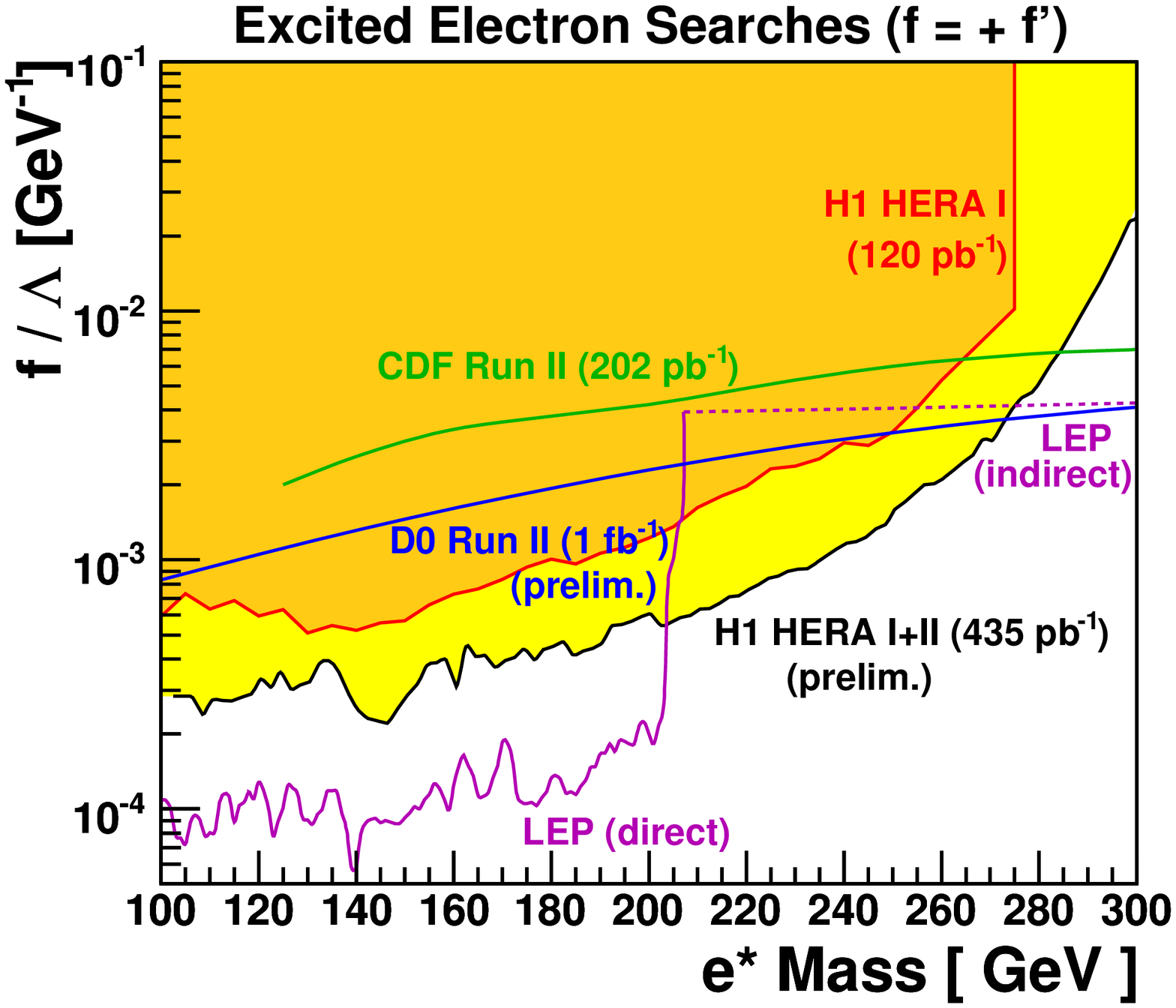}
\caption{\label{fig:exclep} Left: Invariant $ejj$ mass in the H1 
search for $\nu^* \to ejj$. Middle: Invariant $e\gamma$ mass in the $e^* \to e\gamma$ search at D0.
Right: Exclusion limits in the plane of coupling ($f/\Lambda$) and $e^*$ mass. The shaded areas 
and the areas above the lines are excluded. }
\end{figure}

Compositeness models also predict excited states of the familiar fermions denoted e.g. $e^*$, $\nu^*$, 
$q^*$ etc. These decay to a gauge boson and a SM fermion. Searches for these particles have been
carried out at both Tevatron and HERA. Example spectra for the search $\nu^* \to W e \to jj e$ from H1
and $e^* \to e\gamma$ from D0 are shown in Figure~\ref{fig:exclep}. In all cases the data are not showing
any sign for an excess consistent with excited fermion production. A summary of the limits
placed on the mass and coupling of excited electrons is shown in Figure~\ref{fig:exclep}. 
There is a nice competition between the LEP, the Tevatron and the HERA experiments which all exclude 
different and complementary regions of the parameter space.

\section{Conclusions and Outlook}
The current collider experiments continue to vigorously search for new particles as predicted by many 
theories beyond the Standard Model, but so far the data show excellent agreement with the SM expectations
in all experimental signatures. Thus the experiments are placing tighter and tighter limits on the 
mass range allowed for these particles, e.g. in SUSY, Extra Dimension and compositeness models.

There is still more data to come and to be analyzed at the Tevatron and HERA and we look forward
to many more results in the next few years. In Summer 2008 the Large Hadron Collider will 
also start to enter the scene and may finally observe the long-sought new physics at the Tera-scale.

\section*{Acknowledgments}
I wish to thank the organizers of this conference for the excellent preparation and the interesting
and stimulating conference programme.


\section*{References}
\begin{thebibliography}{9}
\bibitem{susy} Wess~J and Zumino~B 1974 {\it Nucl. Phys.} B~\textbf{70} 39

\bibitem{hierarchy1} 
Witten~E., 1981 {\it Nucl. Phys.} B~\textbf{188} 513
\bibitem{hierarchy2} 
Sakai~N, 1981 {\it Z. Phys.} C~\textbf{11} 153
\bibitem{hierarchy3} 
Dimopoulos~S 1981 {\it Nucl. Phys.} B~\textbf{193} 150
\bibitem{unification} Dimopoulos S, Raby A and Wilczek F 1981 {\it Phys. Rev.} D~\textbf{24} 1681
\bibitem{dm1} Ellis~J, Hagelin~J~S, Nanopoulos~D~V, Olive~K and Srednicki~M 1984 {\it Nucl. Phys.} B~\textbf{238} 453
\bibitem{dm2} Goldberg~H 1982 {\it Phys. Rev. Lett}~\textbf{50} 1419
\bibitem{cdfd0sq} Portell X (for the CDF and D0 Collaborations), {\it in these proceedings}
\bibitem{stopll} Abazov V M {\it et al.} (D0 collaboration) submitted to {\it Phys. Lett. B} arXiv:07072846
\bibitem{stopcdf} Aaltonen T {\it et al.} (CDF collaboration) 2007, {\it Phys. Rev.} D~\textbf{76} 072010
\bibitem{stopd0} Abazov V M {\it et al} (D0 Collaboration) 2007 {\it Phys. Lett.} B~\textbf{645} 119;
\bibitem{biscarat} Biscarat C (for the CDF and D0 Collaborations), {\it in these proceedings}
\bibitem{gmsb1} Dine M {\it et al.} 1996 {\it Phys. Rev.} D {\bf 53} 2658
\bibitem{gmsb2} Dine M, Nelson A E and Shirman Y 1995 {\it Phys. Rev.} D{\bf 51} 1362
\bibitem{d0tril} Fox H (for the CDF and D0 Collaborations), {\it in these proceedings}
\bibitem{cdftril} Aaltonen T {\it et al.} (CDF collaboration) 200X, {accepted by Phys. Rev. Lett.}
\bibitem{longlived} Fairbairn M, Kraan A C, Milstead D A, Sj\"ostrand T, Skands P, Sloan T 2007
{\it Phys. Rept.} \textbf{438} 1
\bibitem{splitsusy} Arkani-Hamed N and Dimopoulos S 2005, {\it JHEP} {\bf 06} 073
\bibitem{gmsbstau} Giudice G and Ratazzi R 1999 {\it Phys. Rept.} {\bf 322} 419
\bibitem{champ} Goncharov M (for the CDF and D0 collaboration), {\it in these proceedings}
\bibitem{stopgluino} Abazov V M (D0 Collaboration) 2007 {\it Phys. Rev. Lett.} {\bf 99} 131801
\bibitem{delpho} Aaltonen T {\it et al.} (CDF Collaboration) 2007 {\it Phys. Rev. Lett.} {\bf 99} 1217801
\bibitem{mssmhiggs1} Balasz C, Dias J L, He H J, Tait T and Yuan C P 1999 {\it Phys.
Rev.} D {\bf 59} 055016
\bibitem{mssmhiggs2} Carena M, Mrenna S and Wagner C 1999 {\it Phys. Rev.} D {\bf 60} 075010
\bibitem{mssmhiggs3} Carena M, Mrenna S and Wagner C 200 {\it Phys. Rev.} D {\bf 63} 055008
\bibitem{carena} Balazs C, Carena M and Wagner C 2004 {\it Phys. Rev.} D {\bf 70} 015007
\bibitem{bbb} Jonsson P 2007 {\it in these proceedings}
\bibitem{aurore} Savoy-Navarro A 2007 {\it in these proceedings}
\bibitem{btau} Haba J 2007 {\it in these proceedings}
\bibitem{tau1} Miyazaki Y 2007 {\it in these proceedings}
\bibitem{tau2} Wilson F 2007 {\it in these proceedings}
\bibitem{rarek1} Spadaro T 2007 {\it in these proceedings} KLOE
\bibitem{rarek2} Fandechi R 2007 {\it in these proceedings}
\bibitem{add} Arkani-Hamed N, Dimopoulos S amd Dvali G 1998 {\it Phys. Lett} B\textbf{429} 263.
\bibitem{rs} Randall L and Sundrum R 1999, {\it Phys. Rev. Lett.} \textbf{83} 3370.
\bibitem{zprimecdf} Aaltonen T {\it et al.} (CDF Collaboration) 2007 
{\it submitted to Phys. Rev. Lett.} arXiv:0707.2524
\bibitem{d0rs} Abazov V M {\it et al.} (D0 collaboration) 2007 {\it submitted to Phys. Rev. Lett.} arXiv:0710.3946
\end{thebibliography}
\end{document}